\documentclass[12pt,showpacs,preprintnumbers,amsmath,amssymb]{revtex4}
\usepackage{color} 
\usepackage{graphicx}
\begin{document}

\newcommand{\pderiv}[2]{\frac{\partial #1}{\partial #2}}
\newcommand{\deriv}[2]{\frac{d #1}{d #2}}
\newcommand{\eq}[1]{Eq.~(\ref{#1})}  

\title{The Anisotropic Four-State Clock Model in the Presence of Random
Fields}

\vskip \baselineskip

\author{Octavio D. Rodriguez Salmon}
\thanks{E-mail address:octaviors@gmail.com}
\affiliation{Departamento de F\'{\i}sica \\
Universidade Federal do Amazonas, 3000, Japiim \\
69077-000 \hspace{5mm} Manaus - AM \hspace{5mm} Brazil}

\author{Fernando D. Nobre}
\thanks{Corresponding author: E-mail address: fdnobre@cbpf.br}
\affiliation{Centro Brasileiro de Pesquisas F\'{\i}sicas and \\
National Institute of Science and Technology for Complex Systems \\
Rua Xavier Sigaud 150 \\
22290-180 \hspace{5mm} Rio de Janeiro - RJ \hspace{5mm} Brazil}

\date{\today}


\begin{abstract}
\noindent
A four-state clock ferromagnetic model is studied in the presence of 
different configurations of anisotropies and random fields. 
The model is considered in the limit of infinite-range interactions, 
for which the mean-field approach becomes exact. 
Both representations of Cartesian spin components and two Ising variables 
are used, in terms of which the physical properties and phase diagrams 
are discussed. 
The random fields
follow bimodal probability distributions and the   
richest criticality is found when the fields, applied 
in the two Ising systems, are not correlated.  
The phase diagrams present new interesting 
topologies, with a wide variety of critical points,
which are expected to be useful in 
describing different complex phenomena.  

\vskip \baselineskip



\end{abstract}
\maketitle
\newpage

\section{Introduction}

Spin models represent the most successful applications of 
statistical mechanics and have played  an important role in the 
development of this theory~\cite{lubenskybook,huang}. 
Among those, the Ising model is by far the most investigated,
being able to describe satisfactorily many magnetic materials~\cite{wolfbjp}.
Apart form this, variations of Ising model have been 
considered also for modeling a wide variety 
of systems outside of magnetism, like metallic alloys, lattice 
gases, biological, social, 
and financial systems. 

The introduction of disorder in the Ising model, either in the 
spin-spin interactions (e.g., interactions following a symmetric 
probability distribution, resulting in the Ising spin-glass model), 
and/or by means of a random field acting on 
each spin variable (defining the random-field Ising model), 
has led to further physical realizations,
open problems, and controversial 
aspects~\cite{youngbook,dotsenkobook,nishimoribook}. 
At the infinite-range interaction limit, for which the mean-field 
approach becomes exact, these models have exhibited curious  
properties, and in some cases, very rich critical phenomena that 
has attracted the attention of many 
researchers~\cite{schneiderpytte,aharony,andelman,%
galambirman,mattis,kaufman1,benyoussef,kaufman2,galam95,nogueira98,%
araujo,nuno08a,nuno08b,salmon2009,salmon2010,salmon2014,%
galam87,galam89}).
Even though some properties and criticality found at the mean-field 
level may not occur in more realistic models, defined in terms of 
short-range interactions, random Ising models have been useful 
for investigating several systems, out of the scope of magnetism,     
like neural networks, proteins (particularly, in the study of 
protein folding), optimization problems, and plastic crystals.

The plastic crystals are compounds that present  
an intermediate stable state (called plastic phase) between a 
high-temperature (disordered) liquid phase, and a low-temperature
(ordered) solid phase. 
In such intermediate state, rotational disorder coexists
with translational order, characterized by 
the centers of mass of the molecules forming a regular crystalline 
lattice, with the molecules presenting disorder in their 
orientational degrees of freedom. These systems were treated in the
literature by means of two sets of Ising spin variables representing, 
respectively, the orientational and translational degrees 
of freedom, in addition to a random field acting on one set 
of variables~\cite{galam87,galam89,vives,galam95,salmon2014}. 

Other spin models, mostly defined as generalizations of the 
Ising model, have been much studied in the 
literature~\cite{baxterbook}; among those, one should mention the 
$p$-state Potts model~\cite{wureview}, which appears 
very often in situations where discrete variables with more than 
two states are required for an appropriate description of a given
system. As examples of realizations, one has mixtures of several 
fluids, coloring optimization problems, and monolayers adsorbed on crystal 
surfaces.
Herein, we will be interested in a particular case of the $p$-state 
planar Potts model (also known as clock model), which may be defined in
terms of spin variables $\vec{S}_{i}$, characterized by two
Cartesian components, $\vec{S}_{i} \equiv (S_{ix},S_{iy})$.
Let us introduce a quite general anisotropic Hamiltonian, 

\begin{eqnarray}
{\cal H}(\{h_{ix},h_{iy}\}) = &-& J_{x} \sum_{(ij)}S_{ix} S_{jx}
- J_{y} \sum_{(ij)} S_{iy} S_{jy} - D_{x} \sum_{i=1}^{N} S_{ix}^{2}
- D_{y} \sum_{i=1}^{N} S_{iy}^{2} \nonumber \\ \nonumber \\
\label{eq:clockphamilt}
&-&\sum_{i=1}^{N} h_{ix}S_{ix} -\sum_{i=1}^{N} h_{iy}S_{iy}~, 
\end{eqnarray}

\vskip \baselineskip
\noindent
where  $\sum_{(ij)}$ denote sums over all distinct pairs of 
spins $(i=1,2, \cdots N)$,
$J_{x},J_{y}>0$ are coupling constants, $D_{x},D_{y}>0$ represent 
anisotropy fields, whereas $h_{ix}$ and $h_{iy}$ are random magnetic fields 
acting, respectively, on each spin-variable Cartesian component.
The $p$-state clock variables $\vec{S}_{i}$ are allowed to choose $p$ 
directions in the $xy$ plane, characterized by the components,  

\begin{equation}
\label{eq:pspincomp}
S_{ix} = \cos \theta_{i}~; \qquad 
S_{iy} = \sin \theta_{i}~; \qquad
\theta_{i} = {2 \pi \over p} \, k_{i}~; \qquad 
(k_{i}=0,1,2,\cdots, (p-1))~. 
\end{equation}

\vskip \baselineskip

In the present work we will investigate the case $p=4$ of the 
model above, for which the spin components in~\eq{eq:pspincomp} may be
expressed in terms of two Ising variables. In the next section we
rewrite the Hamiltonian of~\eq{eq:clockphamilt} with these variables, 
discussing 
its properties both in the Cartesian and two-Ising representations; 
in this later case, we derive expressions for the free energy and 
order parameters. In Section III we analyze the phase diagrams of the 
model, by considering bimodal probability distributions for the random
magnetic fields acting on the two sets of Ising variables; the physically
distinct situations of fully correlated and uncorrelated fields 
in these sets of variables are investigated. It is shown that the second case, 
namely, independent probability distributions for each set of 
Ising variables, presents a rich variety of phase diagrams and may
exhibit two distinct ferromagnetic phases, with curious phase
boundaries, ordered critical points, tricritical, and triple points.
We also give heuristic domain-wall arguments
for estimating the lower critical dimension, above which an ordered 
state should appear in the corresponding nearest-neighbor version 
of the model.
Such a rich multicritical behavior is expected to be useful 
for describing magnetic systems, as well as other systems outside 
the area of magnetism, as occurs frequently in many other spin models. 
Finally, in Section IV we present our main conclusions.  

\section{The Two-Ising Representation: Free Energy and Equations of State} 

From now on we will be restricted to the case $p=4$ of the 
model above; hence, the spin components in~\eq{eq:pspincomp} may be
expressed in terms of two Ising variables 
($\tau_{i}= \pm 1$, and $\sigma_{i}= \pm 1$), through the relations 

\begin{equation}
\label{eq:twoisvariables}
S_{ix} = {1 \over 2} \, (\tau_{i} + \sigma_{i}); \quad \quad   
S_{iy} = {1 \over 2} \, (\tau_{i} - \sigma_{i})~.   
\end{equation}

\vskip \baselineskip
\noindent
Considering isotropic coupling constants, i.e.,
$J_{x}=J_{y}$, the Hamiltonian of~\eq{eq:clockphamilt} can be 
rewritten as, 

\begin{equation}
\label{eq:hamiltonian1}
{\cal H}(\{h^{\tau}_{i},h^{\sigma}_{i}\}) = - J \sum_{(ij)}\sigma_{i}\sigma_{j}
- J \sum_{(ij)}\tau_{i}\tau_{j} - D\sum_{i=1}^{N}\tau_{i}\sigma_{i}
-\sum_{i=1}^{N}h^{\tau}_{i}\tau_{i}-\sum_{i=1}^{N}h^{\sigma}_{i}\sigma_{i}~, 
\end{equation}

\vskip \baselineskip
\noindent
where $J>0$ ($J=J_{x}/2=J_{y}/2$) favors ferromagnetic ordering 
in both Ising systems, and the random fields on each set of variables 
are related to those of~\eq{eq:clockphamilt} 
through $h^{\tau}_{i}=(h_{ix}+h_{iy})/2$
and $h^{\sigma}_{i}=(h_{ix}-h_{iy})/2$.  

Comparing Eqs.~(\ref{eq:clockphamilt}) and~(\ref{eq:hamiltonian1}) 
one sees that, through this change of variables, the anisotropy 
fields in the Cartesian-component 
representation result in $D=2(D_{x}-D_{y})$, leading to a coupling 
parameter between the two Ising systems. 
Hence, $D>0$ favors a parallel alignment of  
the spins $\{\tau_{i}\}$ and $\{\sigma_{i}\}$, corresponding 
in~\eq{eq:clockphamilt} to a stronger anisotropy in the $x$-direction, 
whereas the antiparallel alignment of $\{\tau_{i}\}$ and $\{\sigma_{i}\}$
is preferred if $D<0$, a consequence from a larger 
anisotropy field in the $y$-direction. 

The analysis of Ref.~\cite{salmon2014} was inspired on a model for 
plastic crystals~\cite{galam87,galam89,vives,galam95}, defined by means
of two Ising variables, representing respectively, the translational
and rotational degrees of freedom of a molecule. Certainly, these 
variables express very different characteristics of a molecule, and 
particularly, the rotational variables are expected to change more 
freely than the translational ones; 
for this reason, one introduces a random field acting only 
on the rotational degrees of freedom.
In such a model, $h^{\sigma}_{i}=0 \ (\forall i)$ was considered,  
which corresponds in the Hamiltonian
of Eq.~(\ref{eq:clockphamilt}) to $h_{ix}=h_{iy} \ (\forall i)$.
In other systems, e.g., magnetic systems, random fields may result from a  
uniform external field applied in disordered magnetic media, as 
happens for diluted
antiferromagnets~\cite{fishmanaharony,pozenwong,cardy}; in such cases
one should have $h_{ix} \neq h_{iy}$ throughout the material, and  
one expects both random fields 
$h^{\tau}_{i}$  and $h^{\sigma}_{i}$ to play an important role 
in the Hamiltonian of Eq.~(\ref{eq:hamiltonian1}). This 
represents the situation to be considered in the present investigation.

Due to the infinite-range character of the interactions, one can 
write the Hamiltonian of Eq.~(\ref{eq:hamiltonian1}) in the form

\begin{equation}
\label{eq:hamiltonian2}
{\cal H}(\{h^{\tau}_{i},h^{\sigma}_{i}\})= - \frac{J}{2N}{\left 
(\sum_{i=1}^{N}\sigma_{i} \right )}^{2} 
- \frac{J}{2N}{\left (\sum_{i=1}^{N}\tau_{i} \right )}^{2} 
-D\sum_{i=1}^{N}\tau_{i}\sigma_{i} -\sum_{i=1}^{N}h^{\tau}_{i}\tau_{i}
-\sum_{i=1}^{N}h^{\sigma}_{i}\sigma_{i}~,
\end{equation} 

\vskip \baselineskip
\noindent
from which one may calculate the partition function associated with 
a particular realization of the fields $\{ h^{\tau}_{i},h^{\sigma}_{i}\}$, 

\begin{equation}
Z(\{h^{\tau}_{i},h^{\sigma}_{i}\}) =  {\rm Tr} \exp \left[- \beta 
{\cal H}(\{h^{\tau}_{i};h^{\sigma}_{i}\}) \right]~, 
\end{equation}

\vskip \baselineskip
\noindent
where $\beta=1/(kT)$ and 
${\rm Tr} \equiv {\rm Tr}_{\{ \tau_{i},\sigma_{i}=\pm 1 \}}$ indicates a 
sum over all spin configurations. One can now make use of 
the Hubbbard-Stratonovich transformation~\cite{dotsenkobook,nishimoribook}
to linearize the quadratic terms, so that the dependence on the 
index $i$ disappears,

\begin{equation}
Z(\{h^{\tau},h^{\sigma}\}) =  \frac{1}{\pi} \int_{-\infty}^{\infty}dx \, dy \exp(-x^{2}-y^{2}) 
{\left\{ {\rm Tr} \exp [ H(\tau,\sigma,h^{\tau},h^{\sigma})] \right\} }^{N}~,
\end{equation}
 
\vskip \baselineskip
\noindent
where $H(\tau,\sigma,h^{\tau},h^{\sigma})$ is given by

\begin{equation}
H(\tau,\sigma,h^{\tau},h^{\sigma}) = \sqrt{\frac{2\beta J}{N}} \ x \tau + \sqrt{\frac{2\beta J}{N}} \ y \sigma 
- \beta D \tau \sigma + \beta h^{\tau} \tau+ \beta h^{\sigma} \sigma~.   
\end{equation}

\vskip \baselineskip
\noindent
Performing the trace over the spins and defining new variables, related to 
the respective order parameters, 

\begin{equation}
\label{eq:mtausigma}
m_{\tau} = \sqrt{\frac{2kT}{JN}} \ x~; \qquad 
m_{\sigma} = \sqrt{\frac{2kT}{JN}} \ y~, 
\end{equation}

\vskip \baselineskip
\noindent
one obtains

\begin{equation}
Z(\{h^{\tau},h^{\sigma}\})= \frac{\beta J N}{2 \pi} 
\int_{-\infty}^{\infty} dm_{\tau} \, dm_{\sigma} 
\exp[N {\cal G}_{h^{\tau},h^{\sigma}} (m_{\tau},m_{\sigma})]~,   
\end{equation}

\vskip \baselineskip
\noindent
where  

\begin{eqnarray}
{\cal G}_{h^{\tau},h^{\sigma}}(m_{\tau},m_{\sigma}) 
&=& - \frac{1}{2} \beta J m_{\tau}^{2} 
- \frac{1}{2} \beta J m_{\sigma}^{2} + \log \left \{ 
2e^{\beta D} \cosh[\beta J(m_{\tau}+m_{\sigma}+h^{\tau}/J+h^{\sigma}/J)]
\right. \nonumber \\ \nonumber \\
\label{eq:gimtausigma}
&+& \left. 2e^{-\beta D} \cosh[\beta J(m_{\tau}-m_{\sigma}+h^{\tau}/J
-h^{\sigma}/J)] \right \}~.
\end{eqnarray}

\vskip \baselineskip
\noindent
As usual, one considers the thermodynamic limit ($N \rightarrow \infty$), 
and applies the saddle-point method 
to obtain $Z(\{h^{\tau},h^{\sigma}\})$~\cite{dotsenkobook,nishimoribook}. 
So, the free-energy density functional $f(m_{\tau},m_{\sigma})$ results 
from a quenched average of 
$-{\cal G}_{h^{\tau},h^{\sigma}}(m_{\tau},m_{\sigma})$ in~\eq{eq:gimtausigma}, 
over the joint probability distribution  $P(h^{\tau},h^{\sigma})$, 

\begin{equation}
\label{eq:freeenergy}
f(m_{\tau},m_{\sigma}) = \displaystyle  \frac{1}{2} J m_{\tau}^{2} 
+ \frac{1}{2}  J m_{\sigma}^{2}- \frac{1}{\beta}\int_{-\infty}^{\infty}
\int_{-\infty}^{\infty}dh^{\tau}dh^{\sigma}P(h^{\tau},h^{\sigma})
\log Q(h^{\tau},h^{\sigma})~, 
\end{equation}

\vskip \baselineskip
\noindent
with

\begin{eqnarray}
\label{eq:freeenergyq}
Q(h^{\tau},h^{\sigma}) &=& 2e^{\beta D} \cosh[\beta J(m_{\tau}+m_{\sigma} 
+ h^{\tau}/J+h^{\sigma}/J)]  \nonumber \\
&+&2e^{-\beta D} \cosh[\beta J(m_{\tau}-m_{\sigma} + h^{\tau}/J-h^{\sigma}/J)]~. 
\end{eqnarray}

\vskip \baselineskip

If there is no correlation between the 
random fields $\{h^{\tau}_{i}\}$  and $\{h^{\sigma}_{i}\}$, the Hamiltonian
in Eq.~(\ref{eq:hamiltonian1}) presents a symmetry $D \rightarrow -D$, 
together with the inversion of one set of spin variables and its 
associated random field [e.g., $\sigma_{i} \rightarrow -\sigma_{i}$ and 
$h^{\sigma}_{i} \rightarrow -h^{\sigma}_{i}$ $(\forall i)$].   
Since $D=2(D_{x}-D_{y})$ [from Eqs.~(\ref{eq:clockphamilt}) 
and~(\ref{eq:hamiltonian1})], this symmetry corresponds 
to two physically equivalent situations, namely, 
$D_{x}>D_{y} \ (D>0)$ and $D_{x}<D_{y} \ (D<0)$.  
The expression for the free energy above follows this symmetry
[e.g., by considering $D \rightarrow -D$, 
$h^{\sigma} \rightarrow -h^{\sigma}$, and 
$m^{\sigma} \rightarrow -m^{\sigma}$ in~\eq{eq:freeenergyq}]. 

The extremization of the free-energy density above, with respect to the 
parameters $m_{\tau}$ and $m_{\sigma}$, yields the following 
equations of state, 

\begin{eqnarray}
\label{eq:mtau}
m_{\tau} &=& \int_{-\infty}^{\infty}\int_{-\infty}^{\infty}
dh^{\tau}dh^{\sigma}P(h^{\tau},h^{\sigma}) \, 
\frac{R_{+}(h^{\tau},h^{\sigma})}{Q(h^{\tau},h^{\sigma})}~, 
\\ \nonumber \\
\label{eq:msigma}
m_{\sigma} &=& \int_{-\infty}^{\infty}\int_{-\infty}^{\infty}
dh^{\tau}dh^{\sigma}P(h^{\tau},h^{\sigma}) \, 
\frac{R_{-}(h^{\tau},h^{\sigma})}{Q(h^{\tau},h^{\sigma})}~,
\end{eqnarray}

\vskip \baselineskip
\noindent
where 

\begin{eqnarray}
R_{\pm}(h^{\tau},h^{\sigma}) &=& e^{\beta D} \sinh[\beta J(m_{\tau}+m_{\sigma} 
+ h^{\tau}/J+h^{\sigma}/J)] \nonumber \\ 
& \pm & e^{-\beta D} \sinh[\beta J(m_{\tau}-m_{\sigma} 
+h^{\tau}/J-h^{\sigma}/J)]~. 
\end{eqnarray}

\vskip \baselineskip

Now, in order to proceed with the calculations, one has to 
define the joint probability 
distribution $P(h^{\tau},h^{\sigma})$, which appears in 
Eqs.~(\ref{eq:freeenergy}), (\ref{eq:mtau}), and~(\ref{eq:msigma}). 
Herein, we will consider the quite interesting (characterized by 
a rich critical behavior) case of bimodal 
probability distributions~\cite{aharony} in two extreme situations,
namely, fully correlated, and totally uncorrelated fields 
$h^{\tau}$ and $h^{\sigma}$. 

In the first case we will consider $h^{\tau}=h^{\sigma}=h$, 
with $h$ following 

\begin{equation}
\label{eq:hpdf1}
P(h) = \frac{1}{2} \, \delta(h-h_{0}) +\frac{1}{2} \, \delta(h+h_{0})~.  
\end{equation}

\vskip \baselineskip
\noindent
This represents a situation where in each position $i$ the fields 
$\{h^{\tau}_{i}\}$  and $\{h^{\sigma}_{i}\}$ are the same, and  
may be associated to effects due to the randomnesses and 
anisotropies of the 
medium only. In the Cartesian-component representation it 
corresponds to $h_{iy}=0$, so that 
$h^{\tau}_{i}=h^{\sigma}_{i}=h_{ix}/2$ $(\forall i)$. 
Due to this correlation in the random fields, the symmetry 
of the Hamiltonian in Eq.~(\ref{eq:hamiltonian1}),
$D \rightarrow -D$, together with the inversion of one set of 
spin variables and its associated random field, is broken. 
Moreover, since $h_{iy}=0$, this case yields two physically distinct
situations, namely, $D>0 \ (D_{x}> D_{y})$ and $D<0 \ (D_{x}<D_{y})$.

For $D>0$ the system may be described 
by a single order parameter $m$, such that 
$m = m_{\tau} = m_{\sigma}$, leading to the following free energy,

\begin{eqnarray}
\label{eq:dgt0fcorrelated}
f &=& J m^{2}- \frac{1}{ 2 \beta}\log[2 \exp(\beta D)
\cosh[2 \beta J ( m+ h_{0}/J)]
+2 \exp(-\beta D)\cosh[2 \beta J (m+h_{0}/J)]] \nonumber \\ 
&-& \frac{1}{ 2 \beta}\log[2 \exp(\beta D)
\cosh[2 \beta J ( m- h_{0}/J)]+2 \exp(-\beta D)
\cosh[2 \beta J (m-h_{0}/J)]]~,
\end{eqnarray}

\vskip \baselineskip
\noindent
and equation of state, 

\begin{eqnarray}
\label{eq:dgt0mcorrelated}
m &=& \frac{1}{2} \left [ \frac{\sinh[2\beta J(m+h_{0}/J)]}
{\cosh[2\beta J(m+h_{0}/J)]+\exp(-2\beta D)} \right ] \nonumber \\
&+& \frac{1}{2} \left [ \frac{\sinh[2\beta J(m-h_{0}/J)]}
{\cosh[2\beta J(m-h_{0}/J)]+\exp(-2\beta D)} \right ]~.
\end{eqnarray}
 
\vskip \baselineskip
\noindent

On the other hand, for $D<0$ one considers $m = m_{\tau} = -m_{\sigma}$, 
which yields  

\begin{equation}
\label{eq:dlt0fcorrelated}
f= J m^{2}- \frac{1}{\beta}\log[2 \exp(\beta D)
\cosh(2\beta h_{0})+2 \exp(-\beta D)\cosh(2\beta J m)]~,
\end{equation}

\vskip \baselineskip
\noindent
and

\begin{equation}
\label{eq:dlt0mcorrelated}
m = \frac{\sinh(2 \beta J m)}{ \exp(2 \beta D) \cosh(2 \beta h_{0}) 
+ \cosh(2\beta J m)}~. 
\end{equation}

\vskip \baselineskip
\noindent

As a second possibility for the random fields, we take 
$h^{\tau}$ and $h^{\sigma}$ uncorrelated, so the 
joint probability distribution is given by

\begin{equation}
\label{eq:hpdfuncorr}
P(h^{\tau},h^{\sigma}) = P(h^{\tau})P(h^{\sigma})~,
\end{equation}

\vskip \baselineskip
\noindent
and we consider

\begin{equation}
\label{eq:hpdfa}
P(h^{\sigma}) = \frac{1}{2} \, \delta(h^{\sigma}-h_{0}) 
+\frac{1}{2} \, \delta(h^{\sigma}+h_{0})~,   
\end{equation}

\begin{equation}
\label{eq:hpdfb}
P(h^{\tau}) = \frac{1}{2} \, \delta(h^{\tau}-h_{0}) 
+\frac{1}{2} \, \delta(h^{\tau}+h_{0})~,  
\end{equation}

\vskip \baselineskip
\noindent
as the probability distribution functions for the random fields 
acting on each Ising system.

In the Cartesian-component representation this typifies a situation
characterized by local anisotropies, leading to
$h_{ix} \neq h_{iy}$, so that 
both random fields, $h^{\tau}_{i}=(h_{ix}+h_{iy})/2$
and $h^{\sigma}_{i}=(h_{ix}-h_{iy})/2$, play important roles separately.  
These realizations, where in 
each position $i$ one has independent fields, 
$\{h^{\tau}_{i}\}$  and $\{h^{\sigma}_{i}\}$, may result from 
randomnesses of the medium, as well as from other possible
effects (e.g., from the remaining spin variables),  such as to act 
distinctly on the systems $\{\tau_{i}\}$ and $\{\sigma_{i}\}$. 
As mentioned before, this case follows the symmetry 
$D \rightarrow -D$ in~\eq{eq:hamiltonian1}, 
and so an investigation of $D \geq 0$ becomes sufficient. 

After perfoming the integrals in Eqs.~(\ref{eq:freeenergy}), (\ref{eq:mtau}), 
and~(\ref{eq:msigma}), one can show  
that $m_{\tau}=m_{\sigma}$ appears as a solution (due the symmetry of 
the Hamiltonian in~\eq{eq:hamiltonian1}, the 
equivalent solution $m_{\tau}=-m_{\sigma}$ appears in the case 
$D<0$). 
It should be mentioned that, in our numerical analysis, we did not find 
any physically distinct solution from this one for finite 
temperatures; however, as will be shown below, solutions
characterized by $m_{\tau} \neq m_{\sigma}$ appear at $T=0$.
Hence, for $T>0$, similarly to the previous case
(fully correlated random fields), the system will be described 
by a single order parameter $m$, with $m = m_{\tau} = m_{\sigma}$. 
The resulting expressions for the free energy and order parameter are 

\begin{eqnarray}
\label{eq:funcorrelated}
f = Jm^{2}  &-& \frac{1}{4\beta}\log[2\exp(\beta D)\cosh[2\beta J(m+h_{0}/J)]
+2\exp(-\beta D)] \nonumber \\
&-& \frac{1}{4\beta}\log[2\exp(\beta D)\cosh[2\beta J(m-h_{0}/J)]+2\exp(-\beta D)] \nonumber \\ 
&-& \frac{1}{2\beta}\log[2\exp(\beta D)\cosh(2\beta Jm)+2\exp(-\beta D)\cosh(2\beta h_{0})]~,
\end{eqnarray}

\begin{eqnarray}
\label{eq:muncorrelated}
m &=&  \frac{1}{4} \left [  \frac{\sinh[2\beta J(m+h_{0}/J)]}{\cosh[2\beta J(m+h_{0}/J)]
+\exp(-2\beta D)} \right ] \nonumber \\
&+& \frac{1}{4} \left [\frac{\sinh[2\beta J(m-h_{0}/J)]}{\cosh[2\beta J(m-h_{0}/J)]
+\exp(-2\beta D)} \right ] \nonumber \\ 
&+& \frac{1}{2} \left [\frac{\sinh(2\beta Jm)}{\cosh(2\beta Jm)+\exp(-2\beta D)
\cosh(2\beta h_{0})} \right ]~.  
\end{eqnarray}

\vskip \baselineskip
\noindent

In the next section we present and discuss the phase diagrams of the model, 
considering these two choices for the  the joint probability 
distribution $P(h^{\tau},h^{\sigma})$.
In both cases, the equation of state for the single order parameter 
may be expanded as a power series in $m$, 
in the neighborhood of a continuous (second-order) phase transition, 

\begin{equation}
\label{eq:expansion}
m = A_{1}(\beta,D,h_{0})m+A_{3}(\beta,D,h_{0})m^{3}+A_{5}(\beta,D,h_{0})m^{5}+A_{7}(\beta,D,h_{0})m^{7} + \cdots~. 
\end{equation} 

\vskip \baselineskip
\noindent
As usual, the continuous frontiers can be obtained by solving numerically 
the equation $A_{1} = 1$, provided that $A_{3}<0$. In cases where these frontiers end 
at a tricritical point, such a point is obtained by setting $A_{1} = 1$ and $A_{3}=0$, 
conditioned to $A_{5}<0$. 
Furthermore, the so-called fourth-order critical point, after which tricritical 
points do not occur (as a single critical point), 
is located by $A_{1}=1$, $A_{3}=0$, and  $A_{5}=0$, 
provided that $A_{7}<0$. 
The first-order critical frontiers are obtained by standard Maxwell constructions; 
nevertheless, numerical analysis can produce spurious 
solutions, so one must always check if the free energy is minimized.  

All phase diagrams will be represented in terms of dimensionless variables, 
by rescaling conveniently the energy parameters of the system, namely,     
$kT/J$, $h_{0}/J$,  and $D/J$. 
Both ordered ($m \neq 0$) and disordered ($m=0$) 
phases have appeared in our analysis, and as usual, 
they will be labelled by {\rm \bf F} (ferromagnetic) and 
{\rm \bf P} (paramagnetic ) phases, respectively. 
In some of our phase diagrams we find two distinct ferromagnetic
phases (to be labelled by ${\rm \bf F_{1}}$ and ${\rm \bf F_{2}}$), 
separated by a first-order phase transition, characterized
by a jump in their respective magnetizations, $m_{1} \neq m_{2}$. 
It should be emphasized that
the solution $m_{\tau} = m_{\sigma}$ still holds throughout both 
phases ${\rm \bf F_{1}}$ and ${\rm \bf F_{2}}$.   

A wide variety of critical points appeared in our analysis, and herein 
we follow the classification due to Griffiths~\cite{griffiths}: 
(i) a tricritical point signals the encounter of a continuous frontier 
with a first-order line with no change of slope; 
(ii) an ordered critical point corresponds to an isolated critical
point inside the ordered region, terminating a first-order line that
separates two distinct ordered phases;
(iii) a triple point, where three distinct phases coexist, signals the 
encounter of three first-order critical frontiers; 
(iv) a critical end point, where
three phases coexist, corresponding to the intersection of
a continuous line that separates the paramagnetic from one of
the ferromagnetic phases, a first-order line separating the
paramagnetic and the other ferromagnetic phase, and 
a first-order line separating the two ferromagnetic phases;   
(v) a multicritical point, where several phases coexist.
The location
of the critical points defined in (ii)--(v), as well as of the
first-order critical frontiers, were determined by a numerical
analysis of the free-energy minima.
In the phase diagrams we shall use distinct symbols and 
representations for the critical points and frontiers, as described below. 

\begin{itemize}

\item Continuous (second-order) critical frontier: continuous line. 

\item First-order critical frontier: dotted line. 

\item Tricritical point: located by a black circle. 

\item Fourth-order critical point: located by an empty square.

\item Ordered critical point: located by a black asterisk. 

\item Triple point: located by an empty triangle. 

\item Critical end point: located by a black triangle.

\item Multicritical point: located by an empty diamond.
\end{itemize}

These types of behavior appear frequently in many real systems, e.g., 
magnetic compounds and fluid 
mixtures~\cite{lubenskybook,multicrbook,multicrreview,uzunovbook}; 
next, we describe some concrete examples. 
(i) Multicritical phenomena occur along
the surface of magnetic systems,
if the atomic interactions of the surface layer differ significantly from 
those of the bulk~\cite{multicrreview}. In these systems the 
corresponding phase diagrams 
may present distinct ordered phases, as well as regions of coexisting 
ordered-ordered and ordered-disordered states. 
Along these coexistence regions, evidence of 
bicritical, tricritical, and triple points, have been found. 
(ii) A first-order critical line in the plane magnetic field versus temperature, 
terminating at a critical-end point, has been detected in the magnetic
compound ${\rm MnFeAs_{y}P_{1-y}}$, for $y=0.26$~\cite{crendpoint}.  
(iii) Some binary compounds, like ${\rm PrGe_{1.6}}$  and
${\rm CeGe_{1.6}}$ (rare-earth germanides), have shown evidence
of a coexistence of two distinct ferromagnetic phases~\cite{lambert}; 
in particular, the former compound has shown Pr atoms at 
given sites with a substantially larger magnetic moment than those
of Pr atoms at other sites~\cite{papa}. 
The coexisitence of two ferromagnetic phases represents one
of the main results of the present work. 
(iv) Diluted antiferromagnets in the presence of a uniform field 
are considered as physical representations of a ferromagnet 
under random fields~\cite{fishmanaharony,pozenwong,cardy}; 
as a well-known example one could mention the compound 
${\rm Fe_{x}Mg_{1-x}Cl_{2}}$. A curious crossover from a first-order 
to a second-order phase transition has been observed by decreasing
$x$ (this crossover is estimated to happen 
for $x \approx 0.6$)~\cite{kushauerprb96,kushauerjmmm95}. 
One expects such an effect to occur, not as an abrupt change, 
but rather through  
the appearance of some type of multicritical behavior; hence, a fine tunning
of the parameter $x$ in the range $0.6 \leq x \leq 1.0$ is 
highly desirable, and should indicate interesting aspects.  
(v) Fluid mixtures may be, in many cases, mapped into magnetic models, 
in such a way that discrete variables characterized by more than 
two states, like those of the present investigation, 
are necessary for an appropriate description of some ternary and quaternary 
fluid mixtures. These mixtures are good candidates for exhibiting
multicritical phenomena~\cite{lubenskybook,multicrbook,uzunovbook};
it should be mentioned that tricritical points have been observed 
in several multicomponent-fluid mixtures
(a vast list of them may be found in Ref.~\cite{griffithswidom}). 

Before starting a detailed quantitative investigation of the phase 
diagrams and critical points of the model, we first carry out
a ground-state analysis, based 
on the Hamiltonian of~\eq{eq:hamiltonian1} 
[or equivalently, in~\eq{eq:hamiltonian2}]. 
One notices important
competing contributions in the Hamiltonian of~\eq{eq:hamiltonian2}, 
namely, the two quadratic ones, associated with ferromagnetic 
orderings of each system, characterized by a coupling $J$, 
random fields acting on each Ising system with an intensity $h_{0}$, and 
the coupling between the two systems, given by an intensity $|D|$.
In the limit where the random field dominates ($h_{0} \gg J$), 
one expects a disordered state 
({\rm \bf P} phase), where the quadratic terms  
yield zero to the total internal energy $U$.  
However, the resulting energy depends strongly on the coupling $D$,
and particularly in the case of fully correlated random fields, on
the sign of $D$, i.e., through a tendency for aligning the 
two systems parallel to each other if $D>0$, or anti-parallel 
if $D<0$. Hence, for the {\rm \bf P} phase one has 
several possible situations, as described next.  
(a) For uncorrelated random fields, the two systems become 
disordered independently, so that 
$(u/J)= - (2h_{0}/J)$, for both signs of $D$. 
(b) For fully correlated fields and $D>0$, 
each system becomes disordered, but aligned with respect to each other, 
leading to $(u/J)= - (2h_{0}/J)-(D/J)$; however, for $D<0$
each system becomes disordered and anti-parallel 
to each other, so that the random-field contribution
cancels out, leading to $(u/J)= (D/J)$.   
Another important regime concerns the one where the ferromagnetic 
ordering becomes relevant, prevailing over the random-field
contributions ($h_{0} \ll J$). Considering in the present analysis only the 
zero-temperature ferromagnetic ordering with maximum magnetization,  
each quadratic term 
in the Hamiltonian of~\eq{eq:hamiltonian2} contributes with 
$-JN/2$ to the total internal energy. 
Therefore, one obtains 
the internal energy per particle 
$(u/J)= - 1 - (D/J)$, for uncorrelated fields (any $D$), 
as well for fully correlated fields and $D>0$.
However, the case of fully correlated fields and $D<0$ 
has shown to be more subtle, 
deserving a careful quantitive study, as will be discussed in
the next section. 
On the basis of this analysis, one may equate the corresponding 
internal energies to obtain the zero-temperature first-order critical 
frontiers separating the paramagnetic and ferromagnetic phases, for 
fully correlated random fields ($D>0$), as well as
for uncorrelated fields (any $D$). As will be seen in the 
next section, in the first case one
gets the zero-temperature critical field $(h_{0c}/J)=1/2$
(any $D>0)$, whereas in the latter, the critical 
frontier $(D/J)=(2h_{0}/J)-1$ separates the 
phases {\rm \bf P} and ${\rm \bf F_{1}}$.
However, the most interesting and rich 
critical behavior will appear in the case of 
uncorrelated fields, when the parameters
$J$, $D$, and $h_{0}$, 
in the Hamiltonian of~\eq{eq:hamiltonian2}, become
of the same order of magnitude, i.e., 
$(D/J) \approx (h_{0}/J)$, where a multicritical point
emerges; such a zero-temperature point will 
have an important influence on the finite-temperature
phase diagrams.

\section{Results and  Discussions}
\subsection{Correlated Fields}

\begin{figure}[htp]
\begin{center}
\vspace{0.5cm}
\includegraphics[height=5.5cm,angle=0]{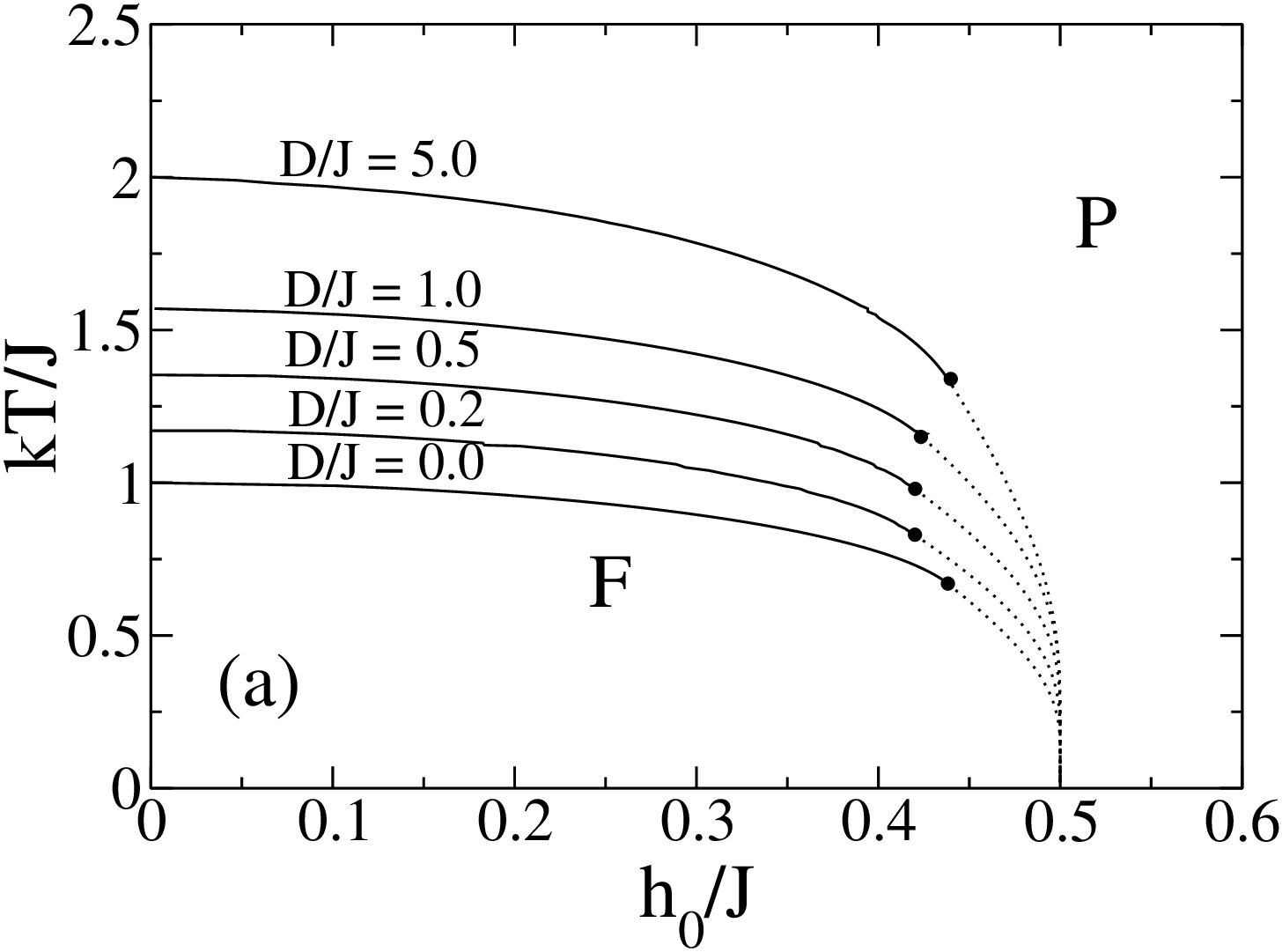}
\hspace{0.3cm}
\includegraphics[height=5.5cm,angle=0]{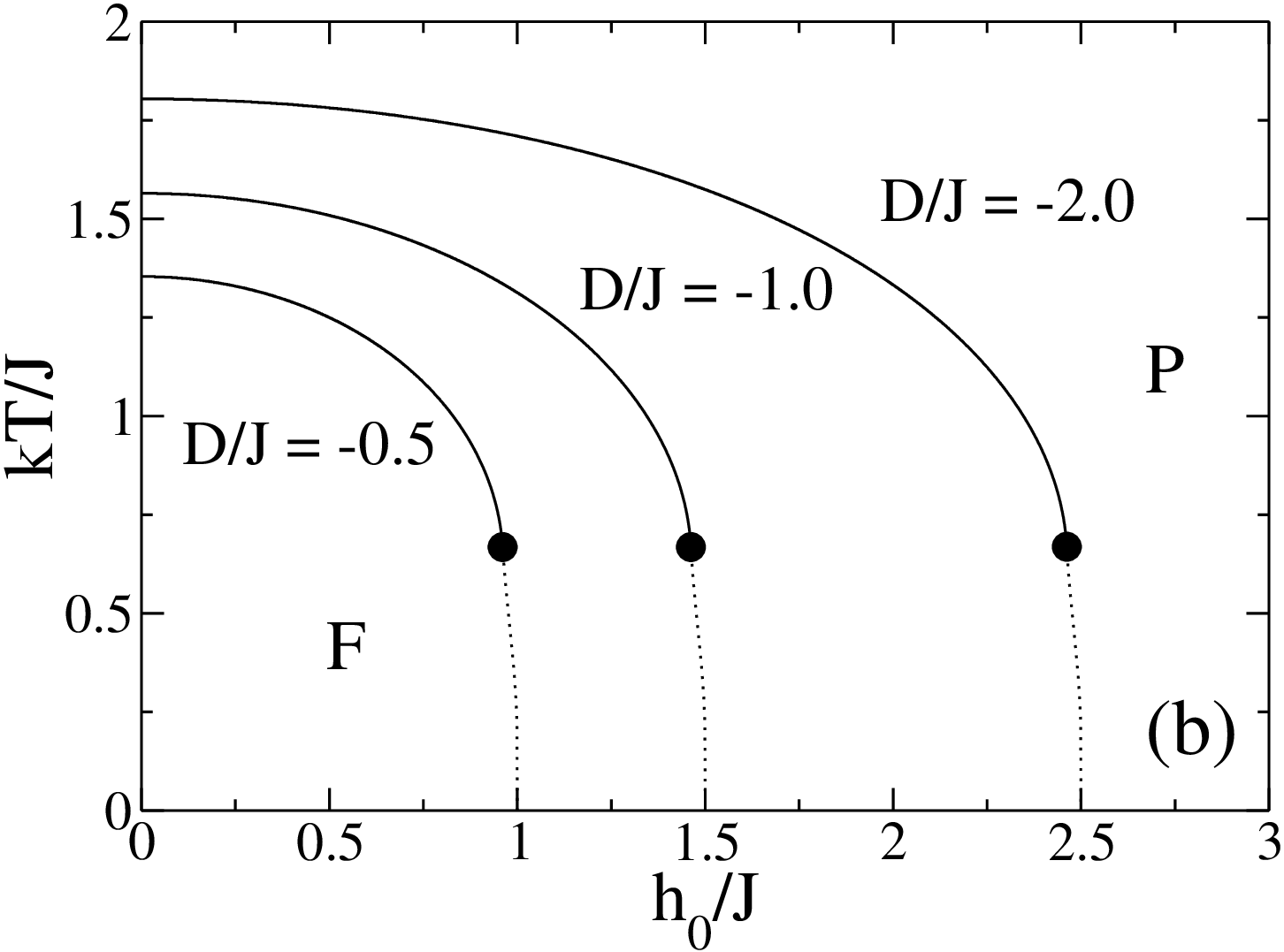}
\end{center}
\vspace{-.5cm}
\caption{
Phase diagram in the plane of conveniently rescaled variables, 
$kT/J$ (dimensionless temperature) versus $h_{0}/J$ (dimensionless
field strength), in the case of fully correlated fields    
$h^{\tau}$ and $h^{\sigma}$. 
(a) Typical values of the dimensionless
coupling $(D/J) \geq 0$;
(b) Typical negative values of the dimensionless
coupling $D/J$.}    
\label{fig:correlatedfields}
\end{figure}

As discussed above, for fully correlated fields 
($h^{\tau}=h^{\sigma}=\pm h_{0}$) one needs to analyze separately
the different signs of the coupling parameter $D$;  
the free energy and order parameter are given, respectively,
by Eqs.~(\ref{eq:dgt0fcorrelated}) and~(\ref{eq:dgt0mcorrelated})
in the case $D>0$, whereas for $D<0$ one should use 
Eqs.~(\ref{eq:dlt0fcorrelated}) and~(\ref{eq:dlt0mcorrelated}). 
The associated phase diagrams are shown in Fig.~\ref{fig:correlatedfields}
in the plane of dimensionless variables $kT/J$ versus $h_{0}/J$.  
From the qualitative point of view, all phase diagrams are similar to the one
of an Ising ferromagnet in the presence of a bimodal random 
field~\cite{aharony}
(which corresponds to the case $D=0$ in Fig.~\ref{fig:correlatedfields}(a), 
i.e., two independent Ising models). 
In analogy to Ref.~\cite{aharony}, 
the two phases {\rm \bf P} and {\rm \bf F} are separated by  
a continuous frontier at high temperatures, followed by a first-order one
for lower temperatures; these two critical lines meet with no change of
slope at a tricritical point (black circle). 

The quantitative differences of the phase diagrams presented in 
Figs.~\ref{fig:correlatedfields}(a) and~\ref{fig:correlatedfields}(b) 
correspond to the enlargement of phase {\rm \bf F} as $|D|$ increases,
characterized by solutions $m>0$, where 
$m = m_{\tau} = m_{\sigma}$ ($D>0$), 
or $m = m_{\tau} = -m_{\sigma}$ ($D<0$). Indeed, 
for $(h_{0}/J)=0$, the critical temperature is 
determined in both cases by solving the equation

\begin{equation}
\label{eq:h0tccorrelated}
{kT_{c} \over J} = \frac{2}{1 + \exp(-2|D|/kT_{c})}~,
\end{equation}

\vskip \baselineskip
\noindent
which comes from setting the coefficient $A_{1}(\beta,D,0)=1$   
in the Landau expansion [cf.~\eq{eq:expansion}] 
for the order parameter given in~\eq{eq:dgt0mcorrelated} ($D>0$),  
or in~\eq{eq:dlt0mcorrelated} ($D<0$). 
In both cases one sees that $(kT_{c}/J) \rightarrow 2$, 
for sufficiently large values of $|D|$, as shown 
in Fig.~\ref{fig:correlatedfields}. 
From~\eq{eq:h0tccorrelated} one notices that the symmetry
$D \rightarrow -D$ is recovered in this particular limit,   
as expected from the Hamiltonian in Eq.~(\ref{eq:hamiltonian1})
in the absence of random fields. 

However, at zero temperature the system is sensitive to the sign of $D$,
and the first-order phase transitions
are obtained by equating the free energies (i.e., internal energies
per particle, $u$)
of the phases {\rm \bf P} and {\rm \bf F}. 
The corresponding critical points $h_{0c}/J$ may be calculated 
analytically either from Eq.~(\ref{eq:dgt0fcorrelated}), 

\begin{equation}
\label{eq:dgt0h0critic}
u_{\rm \bf F} = -(J+D); \quad 
u_{\rm \bf P} = -(2h_{0}+D); \quad \Rightarrow \quad  
{h_{0c} \over J} = {1 \over 2}; \quad (D>0), 
\end{equation}

\vskip \baselineskip
\noindent
or from Eq.~(\ref{eq:dlt0fcorrelated}),

\begin{equation}
\label{eq:dlt0h0critic}
u_{\rm \bf F} = J-(D+2h_{0}); \quad 
u_{\rm \bf P} = D; \quad \Rightarrow \quad  
{h_{0c} \over J} = {1 \over 2} - {D \over J} \quad (D<0).  
\end{equation}

\vskip \baselineskip
\noindent
The zero-temperature critical points of Eqs.~(\ref{eq:dgt0h0critic}) 
and~(\ref{eq:dlt0h0critic}) show a significant difference as one changes 
the sign of the coupling parameter $D$. From Eq.~(\ref{eq:hamiltonian1})
one sees that for $D=0$ one has two independent Ising models, and 
each of them presents a zero-temperature phase transition at 
$(h_{0c}/J)=1/2$, following the Ising ferromagnet in the presence 
of a bimodal random field~\cite{aharony}.
In Fig.~\ref{fig:correlatedfields}(a) this critical point 
remains unchanged by introducing a positive coupling between 
the two Ising systems. 
In the Cartesian-component representation, one should 
remind that the present situation, 
$h^{\sigma}_{i}=h^{\tau}_{i}$, yields $h_{iy}=0$, whereas
$D=2(D_{x}-D_{y})$, so that $D>0$ ($D<0$) corresponds to 
$D_{x}>D_{y}$ ($D_{x}<D_{y}$). Hence, the critical points for 
$D>0$ are associated typically with a random-field phase transition 
in $x$-direction only, and increasing the anisotropy in this direction 
does not change the zero-temperature critical point. 
However, a negative $D$ leads to 
a stronger anisotropy in the $y$-direction, along which there is 
no random field. 
Since the effect of a random field consists in decreasing 
the critical temperature with respect to the one for $h_{0}=0$, 
the preference for the $y$-direction yields a persistence of the {\rm \bf F}
phase for larger values of $h_{0}$, leading to a shift 
of the zero-temperature critical point in 
Fig.~\ref{fig:correlatedfields}(b) according to
$(h_{0c}/J)=1/2-(D/J)$. 
       
\subsection{Uncorrelated Fields}

\begin{figure}[htp]
\begin{center}
\vspace{0.5cm}
\includegraphics[height=5.3cm,angle=0]{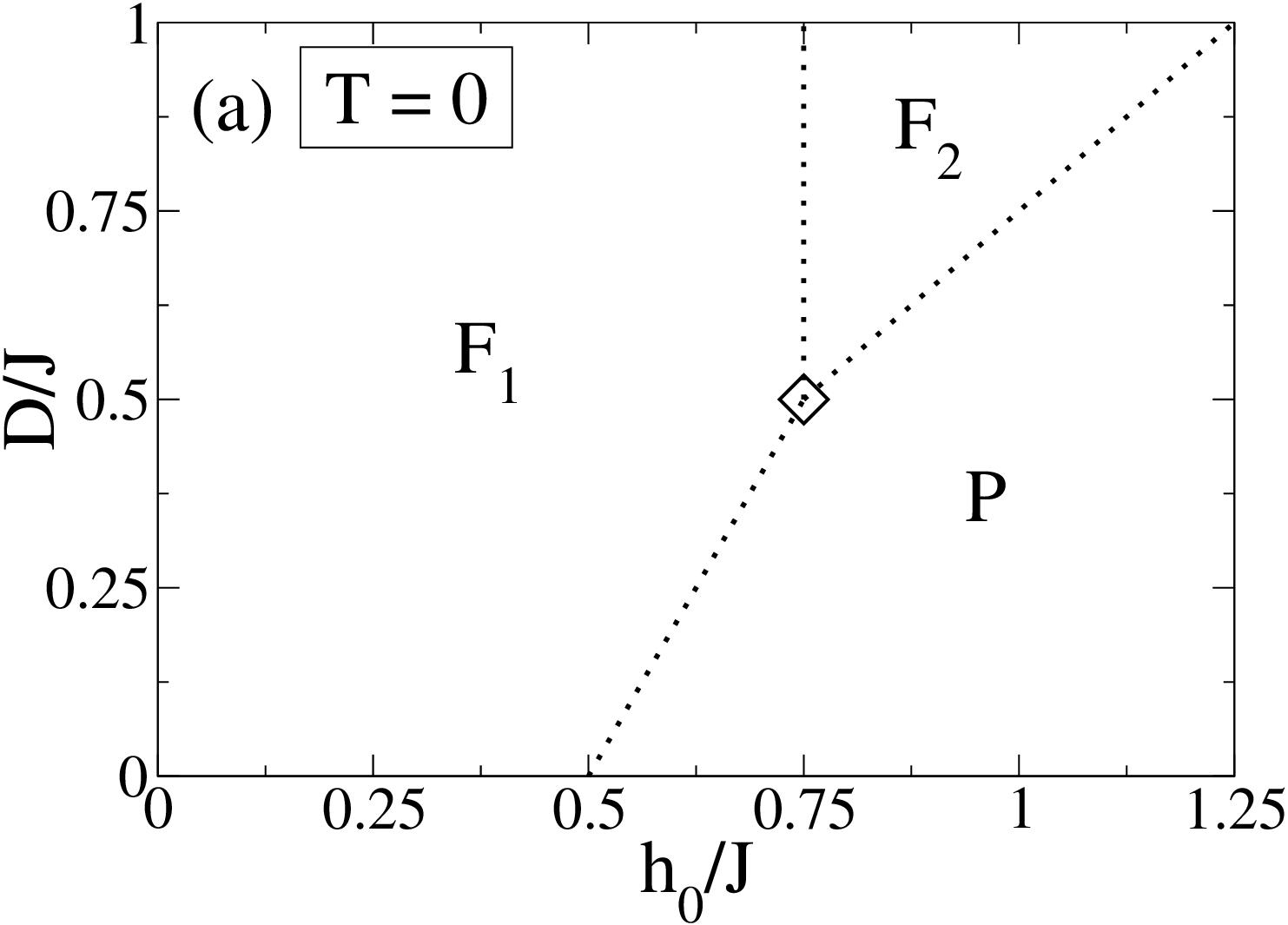}
\hspace{0.3cm}
\includegraphics[height=5.3cm,angle=0]{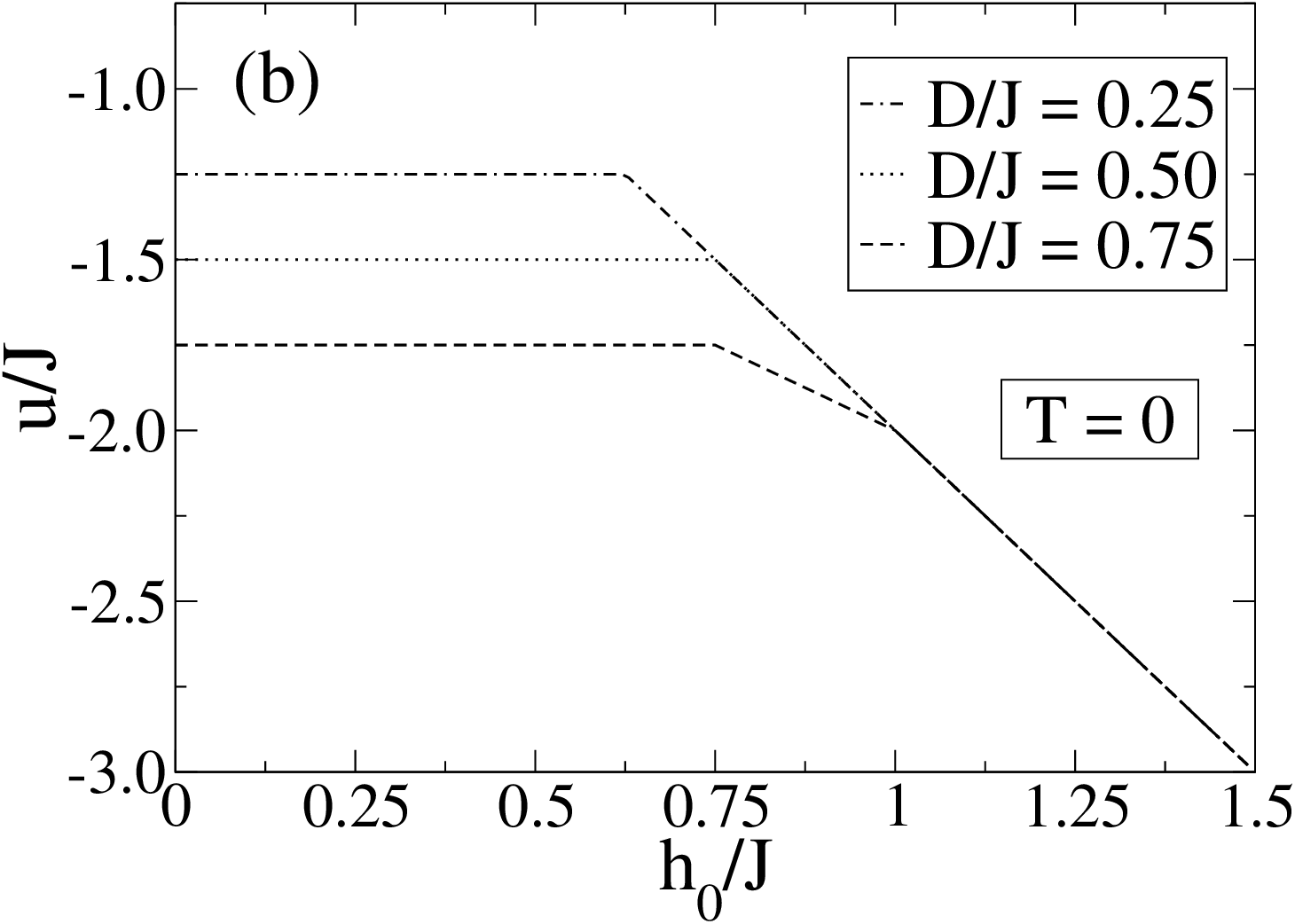}
\end{center}
\vspace{-.5cm}
\caption{
Zero-temperature analysis of the model defined 
in~\eq{eq:hamiltonian1}, in the case of uncorrelated fields
$h^{\tau}_{i}$ and $h^{\sigma}_{i}$. 
(a) Phase diagram in the plane of dimensionless variables
$D/J$ versus $h_{0}/J$; 
all critical frontiers are first order, whereas
at the multicritical point (represented by an empty diamond) one has 
a coexistence of several solutions, as described in the text.
(b) The dimensionless internal energy per particle $u/J$ is shown versus 
$h_{0}/J$ for typical values of $D/J$;  the two limiting values, i.e., 
$(u/J)= - 1 - (D/J)$ (for $h_{0} \ll J$) and 
$(u/J)= - (2h_{0}/J)$  (for $h_{0} \gg J$), predicted in 
the ground-state analysis at the end of the previous section, are verified.}  
\label{fig:zerotemperature}
\end{figure}

According to the discussion of the previous section, 
this case presents the symmetry 
$D \rightarrow -D$ in~\eq{eq:hamiltonian1}, so that from 
now on we restrict ourselves to $D \geq 0$. 
As it will be seen throughout this section, the condition of uncorrelated 
fields leads to a rich criticality, and due to this, we first 
carry out an analysis at zero temperature. 
In Fig.~\ref{fig:zerotemperature}(a) we exhibit
the phase diagram at zero temperature in the plane of dimensionless 
variables $D/J$ versus $h_{0}/J$, 
where three first-order
critical frontiers delimit the phases
{\rm \bf P}, ${\rm \bf F_{1}}$, and  ${\rm \bf F_{2}}$. 

These phases correspond to three
different values of the order parameter $m=m_{\tau}=m_{\sigma}$ that
appear as solutions of~\eq{eq:muncorrelated}, minimizing the Hamiltonian 
given in~\eq{eq:hamiltonian1}: 
$m=0$ (phase {\rm \bf P}), $m=1$ (phase ${\rm \bf F_{1}}$), 
and $m=1/2$ (phase ${\rm \bf F_{2}}$). 
For $0 \leq (D/J) < 1/2$ one has the first two phases only, whereas for 
$(D/J) \geq 1/2$ all three phases become possible. 
In the former case, the 
phases {\rm \bf P} and ${\rm \bf F_{1}}$ are separated 
by a critical frontier given by $(D/J)=(2 h_{0}/J)-1$. 
In the later [$(D/J) \geq 1/2$], one has two first-order frontiers, 
represented by the vertical line $(h_{0}/J)=3/4$ that
separates the ordered phases
${\rm \bf F_{1}}$ and  ${\rm \bf F_{2}}$, and the line
$(D/J)=(h_{0}/J)-1/4$ that divides the ordered phase ${\rm \bf F_{2}}$
from the paramagnetic one.
However, the most interesting aspect of the phase diagram of  
Fig.~\ref{fig:zerotemperature} corresponds to the multicritical point, 
where these three lines meet at $[(h_{0}/J)=0.75,(D/J)=0.5]$ 
(represented by an empty diamond). Curiously, the  order
parameters $m_{\tau}$ and $m_{\sigma}$ yield a coexistence of several 
solutions at this point (some of them breaking the equality 
of these order parameters): 
$(m_{\tau},m_{\sigma})=\{ (-1,-1); (-1/2,-1/2);
(0,0);(0,1/2);(0,-1/2);(-1/2,0);(1/2,0); (1/2,1/2);(1,1) \})$. 
In Fig.~\ref{fig:zerotemperature}(b) we represent 
the dimensionless internal energy per particle $u/J$ versus 
$h_{0}/J$, for typical values of $D/J$ (increasing values of 
$D/J$, from top to bottom). One notices that 
$u/J$ takes a constant value for sufficiently small values of 
$h_{0}/J$ (throughout phase ${\rm \bf F_{1}}$), or decreases linearly 
with $h_{0}/J$ (throughout phases ${\rm \bf F_{2}}$ and {\rm \bf P}), changing
its slope at each critical frontier.  
According to the ground-state analysis at the end of the previous section, 
one has two limiting values for the internal energy, namely,
$(u/J)= - 1 - (D/J)$ (for $h_{0} \ll J$) and 
$(u/J)= - (2h_{0}/J)$ (for $h_{0} \gg J$), which are precisely
those represented in Fig.~\ref{fig:zerotemperature}(b), associated
respectively, with phases ${\rm \bf F_{1}}$ and {\rm \bf P}. Indeed,
by equating these two energies, one obtains the critical frontier 
separating such phases, i.e.,  $(D/J)=(2 h_{0}/J)-1$. 
 
\begin{figure}[htp]
\begin{center}
\vspace{0.5cm}
\includegraphics[height=5.5cm,angle=0]{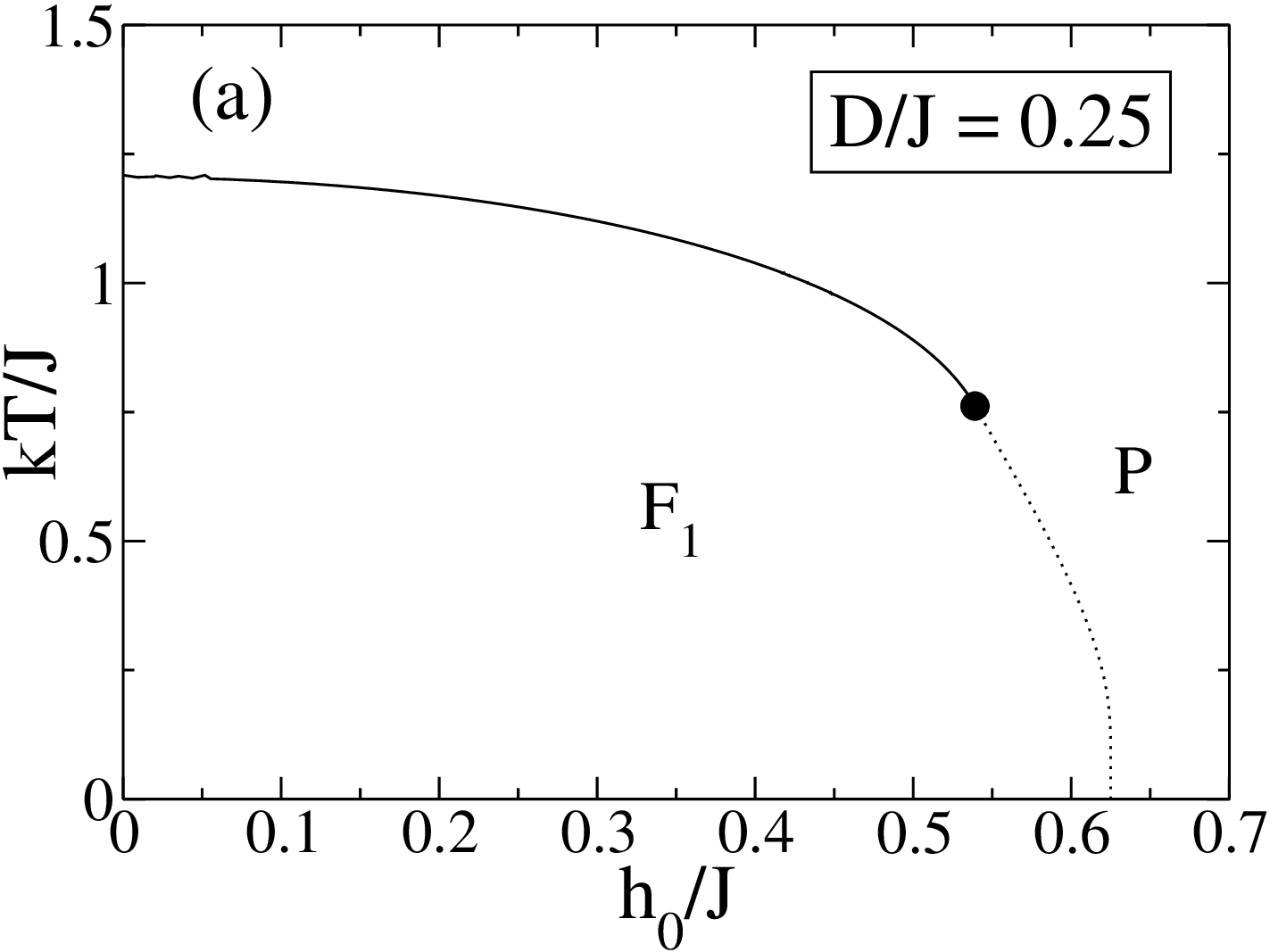}
\hspace{0.3cm}
\includegraphics[height=5.5cm,angle=0]{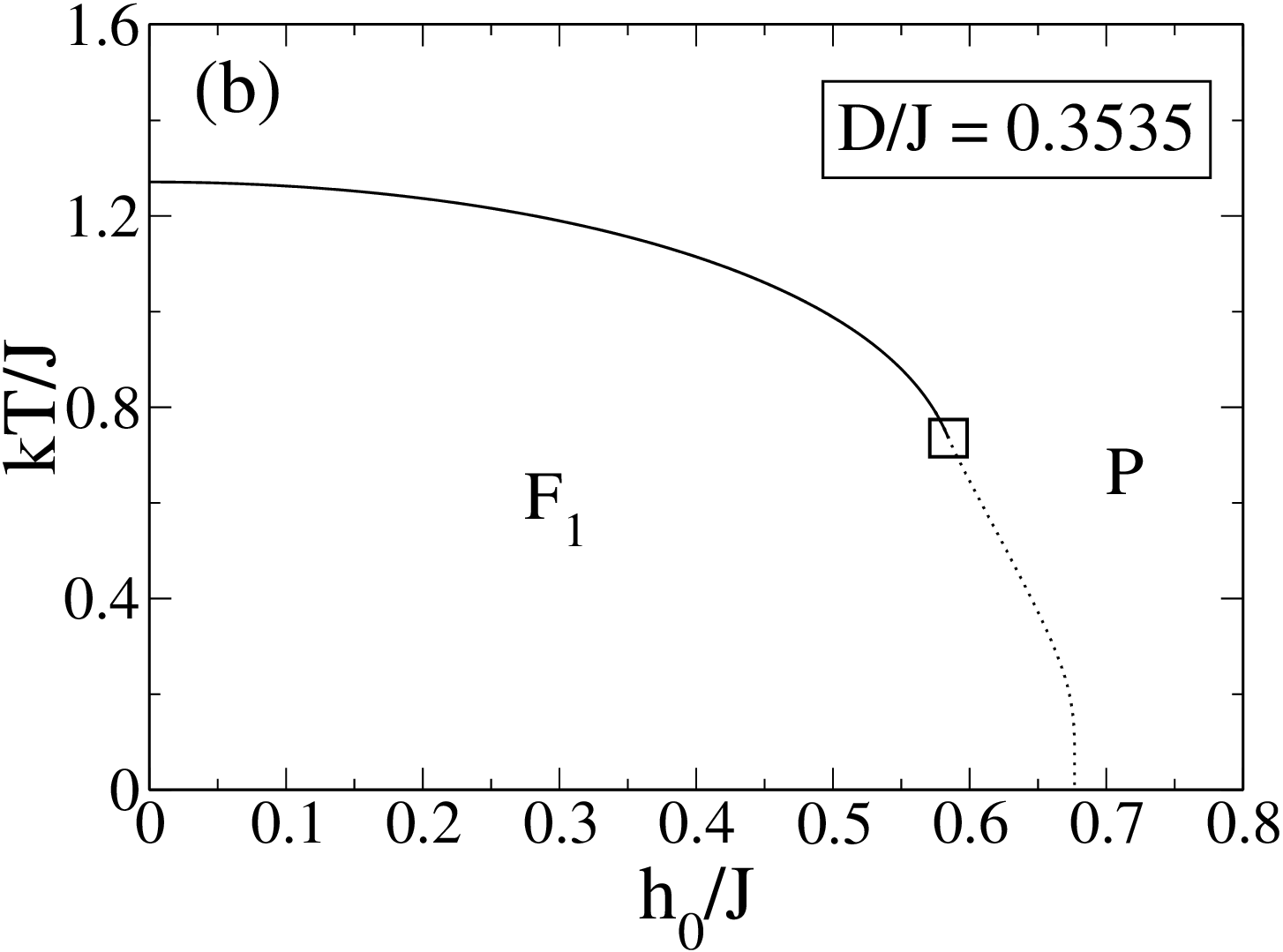}
\end{center}
\vspace{-.5cm}
\caption{
Phase diagrams are exhibited for two typical values of the dimensionless
coupling between the two Ising variables $D/J$, 
in the plane of conveniently rescaled variables, 
$kT/J$ (dimensionless temperature) versus $h_{0}/J$ (dimensionless
field strength). 
(a) Case $(D/J)=0.25$, showing a tricritical point (black circle), 
where a continuous frontier (high temperatures) meets a first-order
critical frontier (low temperatures); we shall refer
to this type of phase diagram as topology I. 
(b) Case $(D/J)=0.3535$, where the empty square denotes a 
fourth-order critical point, which represents the limit for the 
appearance of the single tricritical point shown in (a) (see text).}  
\label{fig:tricritical}
\end{figure}

In Fig.~\ref{fig:tricritical} we present phase diagrams 
for two typical values of the dimensionless coupling
between the two sets of Ising variables, namely, $(D/J)=0.25$
and $(D/J)=0.3535$, with critical frontiers separating the ferromagnetic phase 
${\rm \bf F_{1}}$ (sufficiently small values of $kT/J$ and $h_{0}/J$) from
the paramagnetic phase {\rm \bf P}. 
In Fig.~\ref{fig:tricritical}(a) one notices that 
the value of the coupling $D/J$ is not sufficiently strong to 
change qualitatively the phase diagram of an Ising ferromagnet in the 
presence of a bimodal random field~\cite{aharony}, where one 
finds a tricritical point signalling the encounter 
of the continuous frontier (high temperatures) with a first-order
critical frontier (low temperatures). In this case, the only 
quantitative effect concerns an enlargement of phase ${\rm \bf F_{1}}$
by increasing $D/J$; such a phase diagram will 
be referred from now on as topology I, and it appears 
for $0<(D/J)<0.3535$. Such a topology ends up at
$(D/J)=0.3535$, where the tricritical point turns into a
fourth-order critical point [located at $(h_{0}/J)=0.585; (kT/J)=0.735$ and 
represented by the empty square in Fig.~\ref{fig:tricritical}(b)]. 
Fourth-order critical points
were found in other disordered spin models, like those treated in 
references~\cite{mattis,kaufman1,salmon2009,salmon2010}; 
they are sometimes entitled in the literature as ``vestigial'' tricritical
points, because they delimit the existence of those critical
points~\cite{kaufman1}.  

\begin{figure}[htp]
\begin{center}
\vspace{0.5cm}
\includegraphics[height=5.5cm,angle=0]{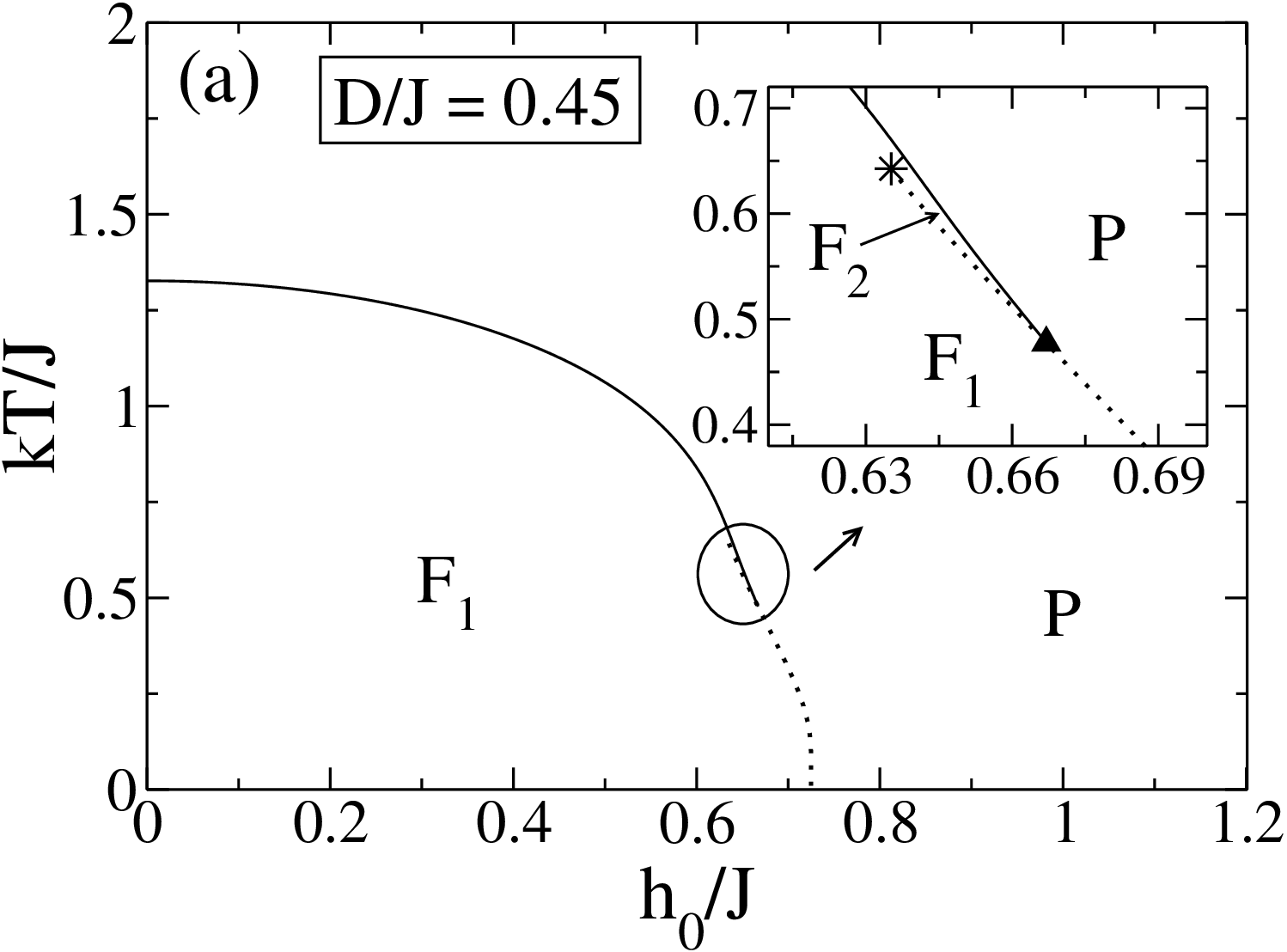}
\hspace{0.3cm}
\includegraphics[height=5.5cm,angle=0]{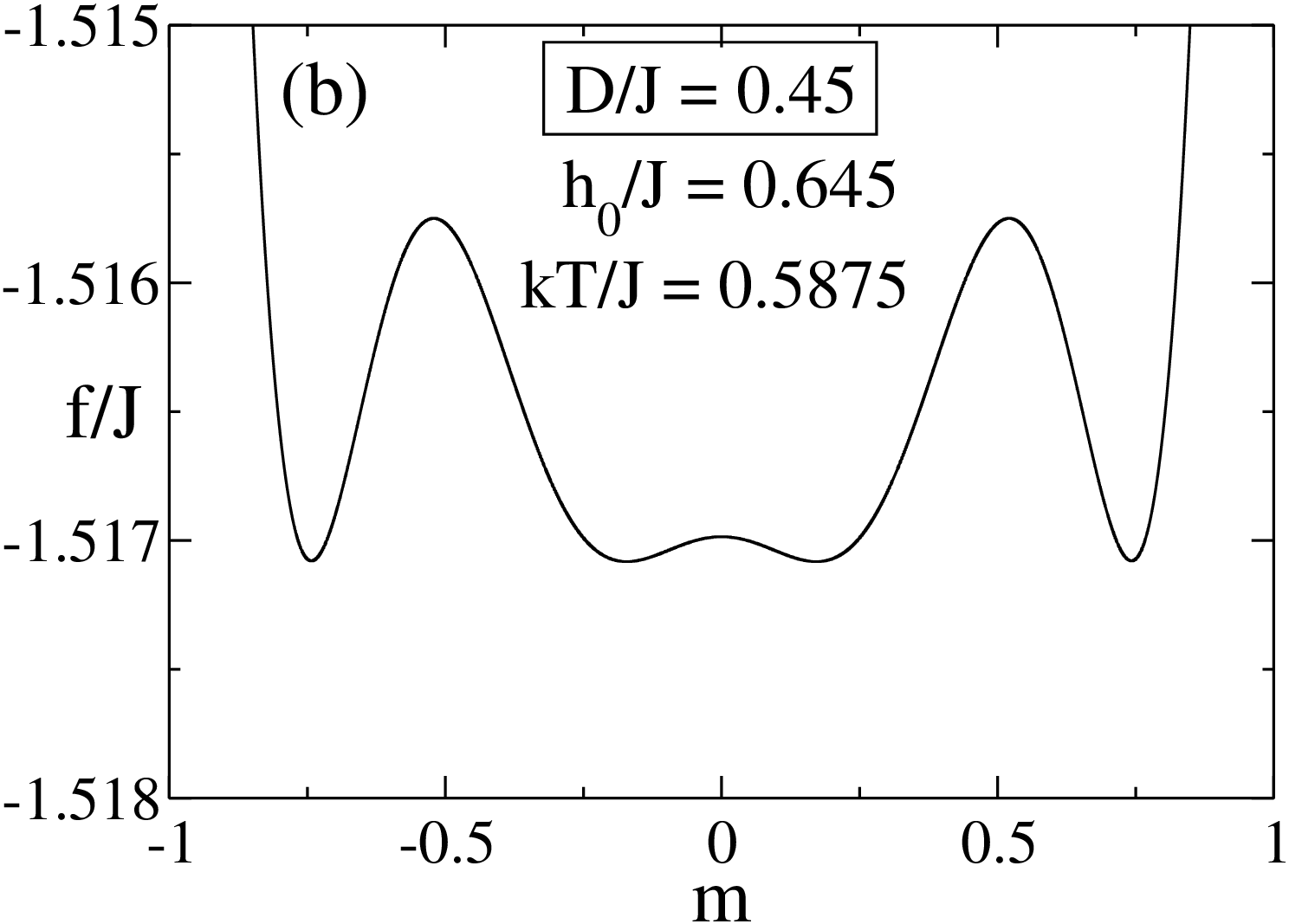}
\end{center}
\vspace{-.5cm}
\caption{
(a) Phase diagram for a typical value of the dimensionless
coupling between the two Ising variables, $(D/J)=0.45$, 
in the plane of conveniently rescaled variables, 
$kT/J$ (dimensionless temperature) versus $h_{0}/J$ (dimensionless
field strength). The ordered phase ${\rm \bf F_{2}}$ appears
in a small part of the phase diagram, as shown in the enlargement
of the inset; the black triangle and the asterisk represent  
a critical end point and an ordered 
critical point, respectively; we shall refer
to this type of phase diagram as topology II.    
(b) The dimensionless free energy is plotted versus the dimensionless order
parameter, for a point of the phase diagram located at  
$[(h_{0}/J)=0.6450; (kT/J)=0.5875]$, belonging to the first-order frontier 
shown in the inset of (a), which divides the phases ${\rm \bf F_{1}}$ 
and ${\rm \bf F_{2}}$.} 
\label{fig:secondtopology}
\end{figure}

For $(D/J)>0.3535$ the additional ordered phase ${\rm \bf F_{2}}$
arises, although for a certain range of values of $D/J$  
it may occupy a small part of the phase diagram, as 
shown in Fig.~\ref{fig:secondtopology}(a) for the case $(D/J)=0.45$. 
In the inset of Fig.~\ref{fig:secondtopology}(a) one sees the piece of the 
first-order critical frontier that separates the phases ${\rm \bf F_{1}}$ 
and ${\rm \bf F_{2}}$, delimiting phase ${\rm \bf F_{2}}$,  
from the critical end point
(represented by a black triangle) to the ordered critical point (represented
by an asterisk). From now on, we shall refer
to this type of phase diagram as topology II, and as it will
be discussed next, this topology applies for      
$0.3535<(D/J)<0.470$. 
In this case, the border of the {\rm \bf P} phase is given by a continuous 
part (high temperatures) that ends up at a critical end point, being
followed by a first-order critical frontier (low temperatures). For temperatures
right above the critical end point, one can go continuously from the {\rm \bf P}         
phase to ${\rm \bf F_{2}}$; however, most of the {\rm \bf P} border
is shared with the ${\rm \bf F_{1}}$ phase, as shown 
in Fig.~\ref{fig:secondtopology}(a). In Fig.~\ref{fig:secondtopology}(b) we 
plot the dimensionless free energy versus the dimensionless order
parameter, for a point of the phase diagram belonging to the first-order frontier 
shown in the inset of (a), dividing phases ${\rm \bf F_{1}}$ 
and ${\rm \bf F_{2}}$; one sees clearly the 
coexistence of two different values of $|m|$, typical of a first-order criticality.
 
\begin{figure}[htp]
\begin{center}
\vspace{0.5cm}
\includegraphics[height=6.0cm,angle=0]{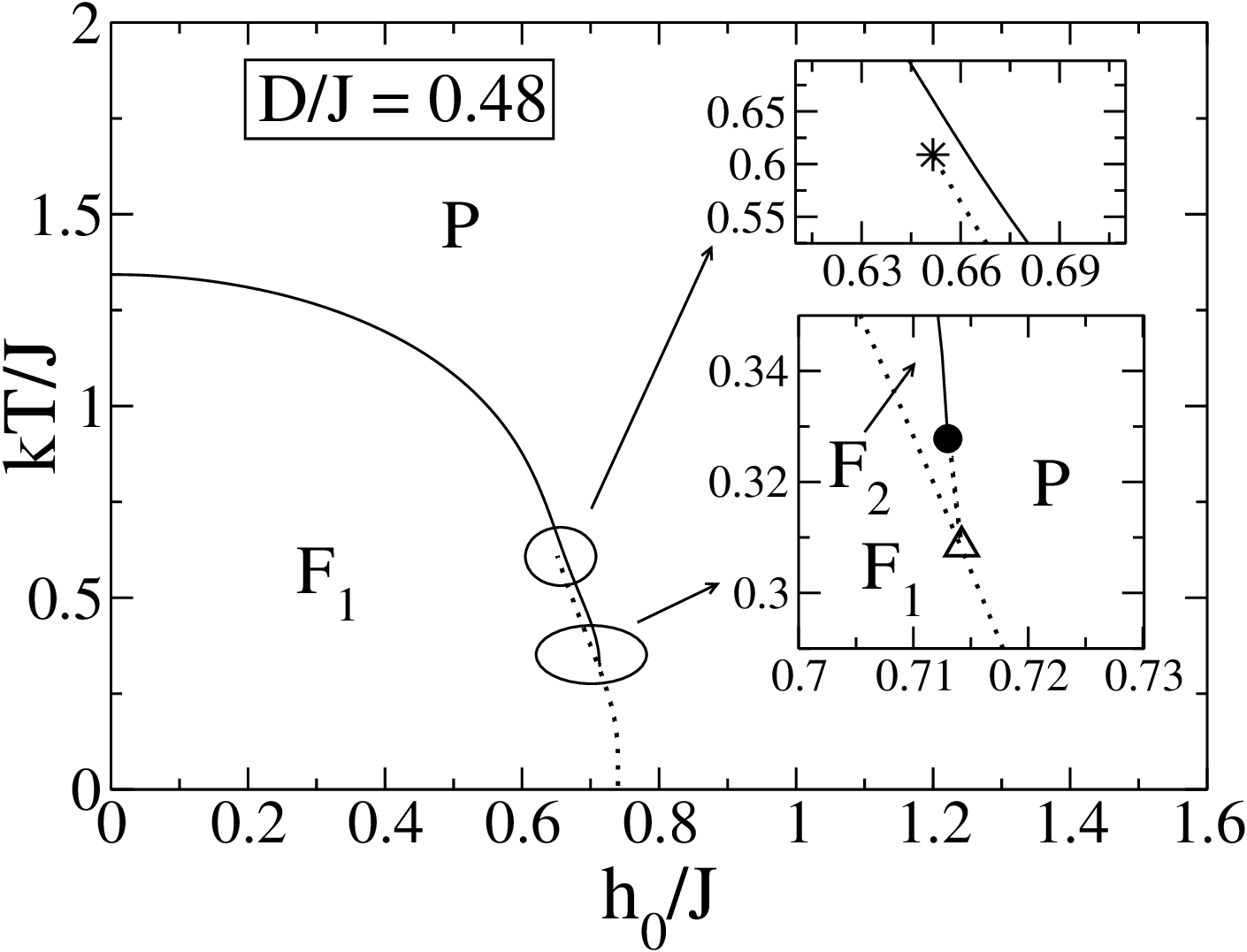}
\end{center}
\vspace{-.5cm}
\caption{
Phase diagram for a typical value of the dimensionless
coupling between the two Ising variables, $(D/J)=0.48$, 
in the plane of conveniently rescaled variables, 
$kT/J$ (dimensionless temperature) versus $h_{0}/J$ (dimensionless
field strength). The ordered phase ${\rm \bf F_{2}}$ and critical points
are shown in the insets, where we have enlarged two important 
regions of the phase diagram, represented by ellipses. In the lower 
inset one sees a triple point (empty triangle) and a tricritical point (black 
circle), which emerged from the critical end point 
of Fig.~\ref{fig:secondtopology}(a). In the upper inset we show the ordered
critical point signalling the end of phase ${\rm \bf F_{2}}$.  
We shall refer to this type of phase diagram as topology III.}    
\label{fig:thirdtopology}
\end{figure}

By increasing gradually $D/J$ we have verified that 
the critical end point of Fig.~\ref{fig:secondtopology}(a) disappears, 
giving rise 
to two other critical points, namely, a triple and a tricritical one.  
This is shown in Fig.~\ref{fig:thirdtopology} where we present the 
phase diagram for $(D/J)=0.48$; the ordered phase ${\rm \bf F_{2}}$, 
as well as the critical points
are shown in the insets, through enlargements of two relevant parts 
of the critical region. 
In order to determine the upper limit associated with topology II, we had 
to estimate numerically the value of $D/J$ for which the tricritical point emerges, 
leading to topology III. We have found that this occurs for $(D/J) = 0.471 \pm 0.001$, 
in the sense that topology II holds clearly for $(D/J)=0.470$, whereas topology
III applies for $(D/J)=0.472$.  
In this later topology, the border of the {\rm \bf P} phase presents a rather 
rich critical behavior, whereas the critical frontier between the two ordered phases 
(${\rm \bf F_{1}}$ and ${\rm \bf F_{2}}$) is first-order, terminating in an
ordered critical point, similarly to the one shown in Fig.~\ref{fig:secondtopology}(a). 
The border of the {\rm \bf P} phase is composed by a continuous 
part (high temperatures) that ends up at a tricritical point, being
followed by a small first-order critical frontier down to the triple point, 
below which a first-order phase transition separates phases {\rm \bf P}         
and ${\rm \bf F_{1}}$. The frontier between phases {\rm \bf P}         
and ${\rm \bf F_{2}}$ is either continuous (above the tricritical point), or
first-order (between the tricritical and triple points).  
The region of the two insets of Fig.~\ref{fig:thirdtopology}
suggests that such a rich critical behavior should be influenced by the 
zero-temperature multicritical point (located at $[(h_{0}/J)=0.75,(D/J)=0.5]$
in Fig.~\ref{fig:zerotemperature}), where one has a coexistence of nine
different solutions for the order parameters. 

\begin{figure}[htp]
\begin{center}
\vspace{0.5cm}
\includegraphics[height=5.5cm,angle=0]{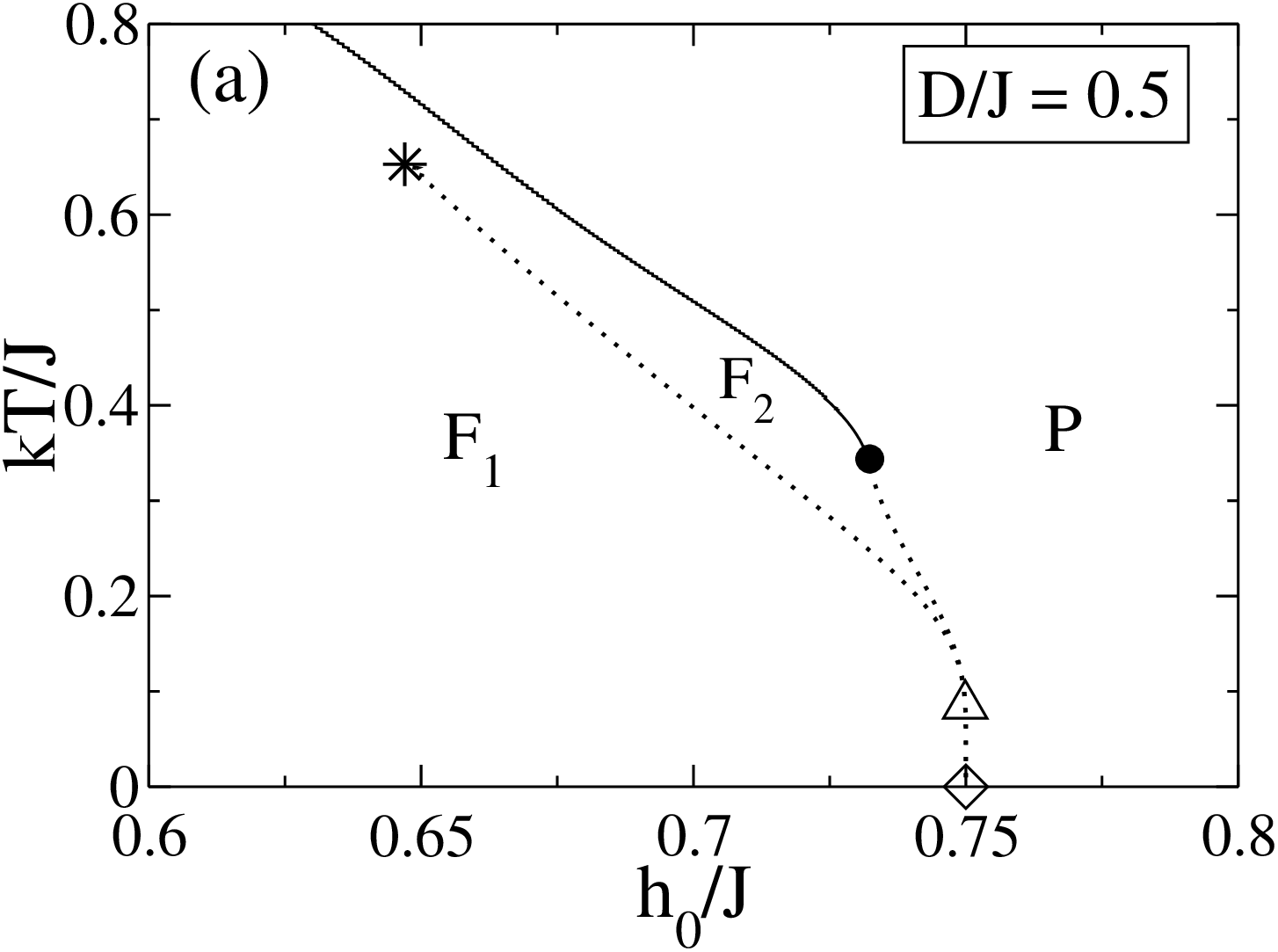}
\hspace{0.3cm}
\includegraphics[height=5.5cm,angle=0]{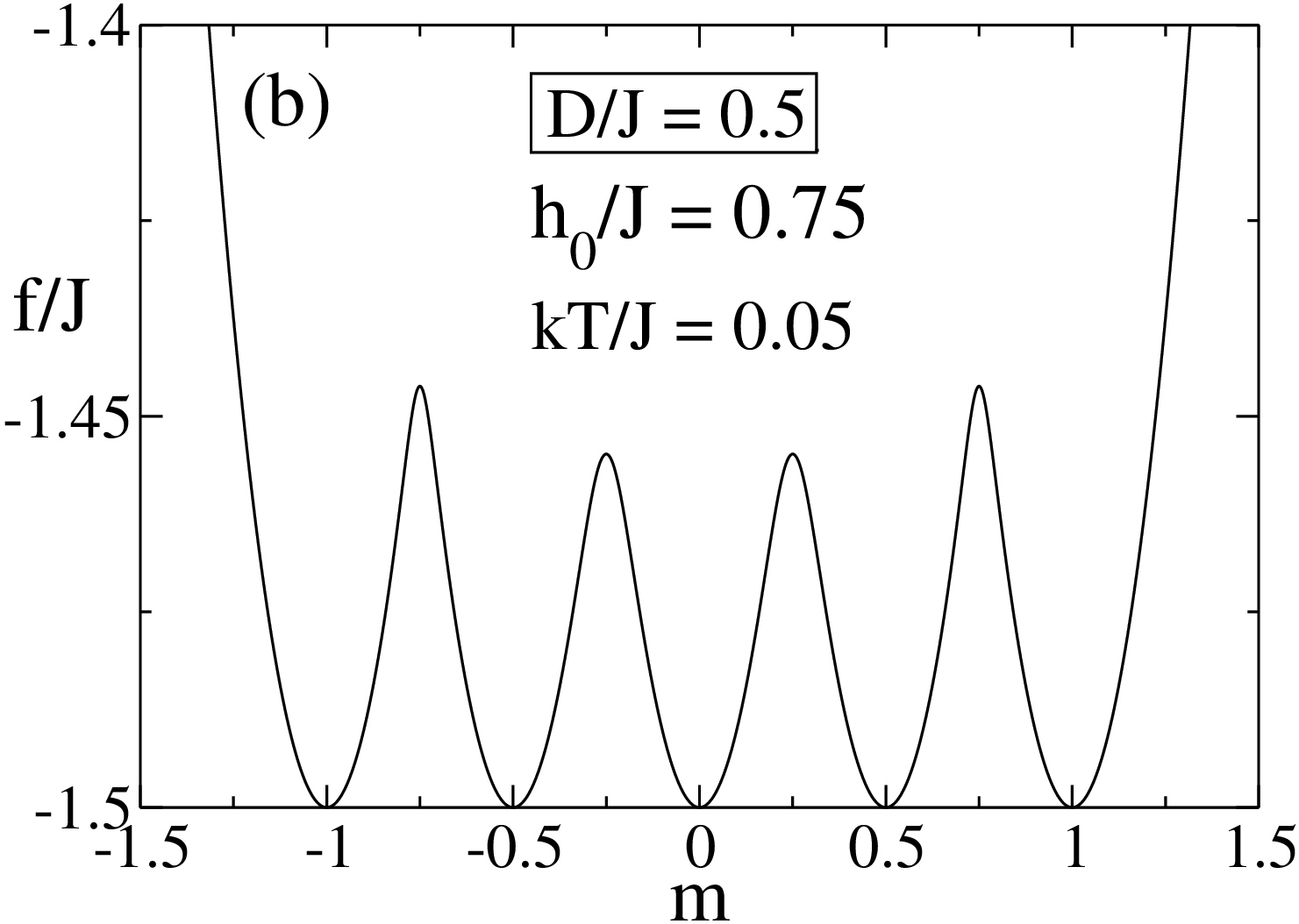}
\end{center}
\vspace{-.5cm}
\caption{
(a) Phase diagram for the dimensionless
coupling $(D/J)=0.5$, 
in the plane of conveniently rescaled variables, 
$kT/J$ (dimensionless temperature) versus $h_{0}/J$ (dimensionless
field strength). The two ordered phases (${\rm \bf F_{1}}$ and ${\rm \bf F_{2}}$)
are separated by a first-order critical frontier that terminates at an ordered critical point
(represented by an asterisk). The border of the {\rm \bf P} phase presents a 
tricritical (black circle) and a triple point (empty triangle) at finite temperatures,  
whereas at zero temperature one finds the multicritical point of 
Fig.~\ref{fig:zerotemperature} (represented by an empty diamond).
We shall refer to this type of phase diagram as topology IV. 
(b) The dimensionless free energy is plotted versus the dimensionless order
parameter, for a point of the phase diagram located at  
$[(h_{0}/J)=0.75; (kT/J)=0.05]$, belonging to the low-temperature
first-order frontier delimited by the triple and multicritical points.}
\label{fig:fourthtopology}
\end{figure}

For $(D/J)=0.5$, topology III ends up through the  appearance of the multicritical 
point at zero temperature, as shown  in Fig.~\ref{fig:fourthtopology}(a) (to be 
referred hereafter as topology IV).  
This point (represented by the empty diamond) corresponds to the 
multicritical point already 
exhibited in Fig.~\ref{fig:zerotemperature}, and, as expected, it occurs only
for $(D/J)=0.5$. In this sense, topology III applies for $0.472\leq(D/J)<0.5$, 
whereas topology IV holds only for $(D/J)=0.5$. 
Comparing Figs.~\ref{fig:thirdtopology} and~\ref{fig:fourthtopology}(a) one 
notices, besides the zero-temperature multicritical point, an enlargement of 
phase ${\rm \bf F_{2}}$, essentially due to fact that the triple point is now 
located at a much lower temperature, maintaining the topological 
structure shown in the insets of Fig.~\ref{fig:thirdtopology}. 
To illustrate the low-temperature critical behavior, in Fig.~\ref{fig:fourthtopology}(b) 
we plot the dimensionless free energy versus the dimensionless order
parameter, for a point of the phase diagram belonging to the
first-order frontier delimited by the triple and the multicritical points. 
There, the free energy exhibits solutions corresponding to the 
two ordered phases (${\rm \bf F_{1}}$ and ${\rm \bf F_{2}}$) 
coexisting with the disordered phase one ($m=0$). We verified that when
this first-order frontier approaches zero temperature (close to the  multicritical point), 
four local minima, characterized by higher values of $f/J$, corresponding to 
$(m_{\tau},m_{\sigma})=\{(0,1/2);(0,-1/2);(-1/2,0);(1/2,0) \}$, 
approach the five coexisting global minima shown in  
Fig.~\ref{fig:fourthtopology}(b), corresponding to 
$(m_{\tau}=m,m_{\sigma}=m)=\{ (-1,-1); (-1/2,-1/2); (0,0);(1/2,1/2);(1,1) \}$. 
Accordingly, nine phases will coexist when the lower first-order curve touches  
the multicritical point at zero temperature. 
Therefore, this corresponds to the only point at which one finds solutions  
with $m_{\tau} \neq m_{\sigma}$, as discussed in the zero-temperature
phase diagram of Fig.~\ref{fig:zerotemperature}.

\begin{figure}[htp]
\begin{center}
\vspace{0.5cm}
\includegraphics[height=5.5cm,angle=0]{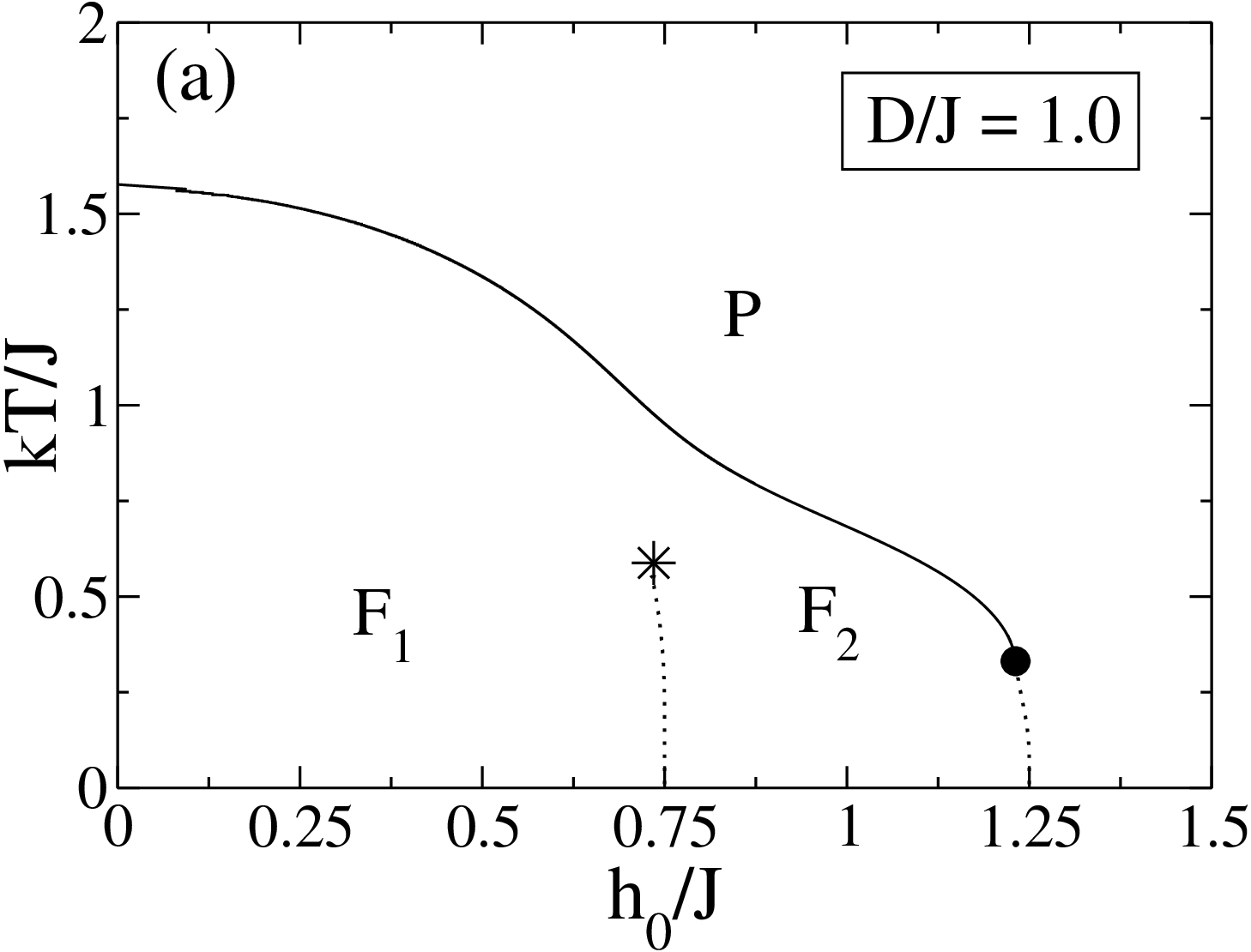}
\hspace{0.2cm}
\includegraphics[height=5.5cm,angle=0]{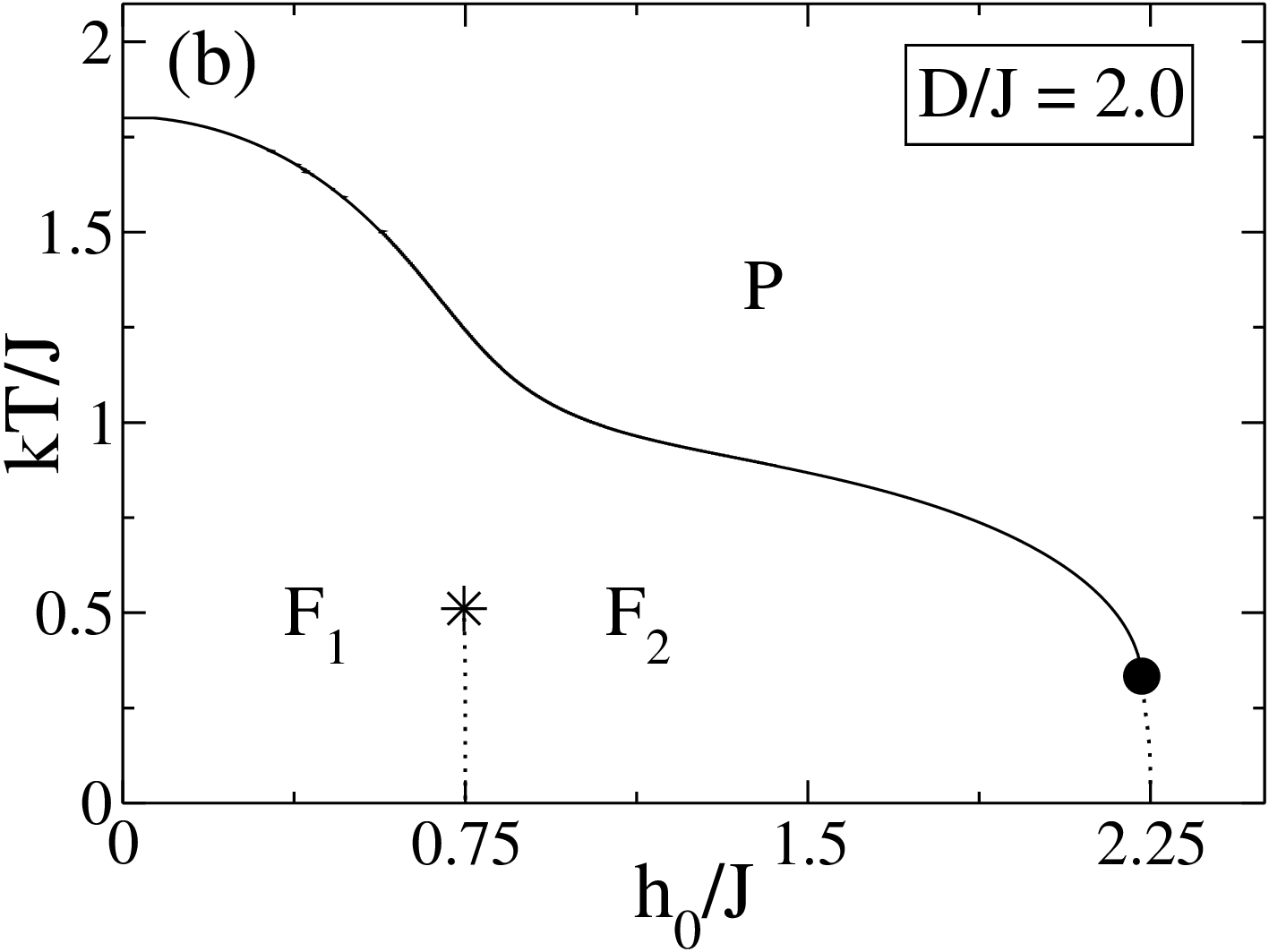}
\end{center}
\vspace{-.5cm}
\caption{ 
Phase diagrams for two typical values of the dimensionless
coupling $D/J$ [with $(D/J)>0.5$], 
in the plane of conveniently rescaled variables, 
$kT/J$ (dimensionless temperature) versus $h_{0}/J$ (dimensionless
field strength). The two ordered phases (${\rm \bf F_{1}}$ and ${\rm \bf F_{2}}$)
are separated by a first-order critical frontier that terminates at an ordered critical point
(represented by an asterisk). The border of the {\rm \bf P} phase presents a 
tricritical point only (black circle). 
(a) Case $(D/J)=1.0$; (b) Case $(D/J)=2.0$; we shall refer to this type of phase 
diagram as topology V. In these cases, according to the phase diagram 
at zero-temperature
(cf. Fig.~\ref{fig:zerotemperature}), the first-order frontier separating
phases ${\rm \bf F_{1}}$ and ${\rm \bf F_{2}}$ starts at  
$(h_{0}/J)=0.75$, whereas the one dividing phases 
${\rm \bf F_{2}}$ and {\rm \bf P} starts at
$(h_{0}/J)=(D/J) +1/4$.}    
\label{fig:fifthtopology}
\end{figure}

In Fig.~\ref{fig:fifthtopology} we exhibit phase diagrams
for two typical values of the dimensionless
coupling $D/J$ [with $(D/J)>0.5$], corresponding to topology V. 
In contrast to topologies II--IV, one sees clearly that the first-order frontier 
dividing phases ${\rm \bf F_{1}}$ and ${\rm \bf F_{2}}$ appears now 
shifted from the one that divides ${\rm \bf F_{2}}$ and {\rm \bf P}. 
 This aspect has to do with the zero-temperature phase diagram presented 
in Fig.~\ref{fig:zerotemperature}, where these two frontiers for $T=0$
start, respectively, at $(h_{0}/J)=0.75$ and 
$(h_{0}/J)=(D/J) +1/4$, for $(D/J)>0.5$.  
The border of the {\rm \bf P} phase is now characterized by a change
of concavity, as well as by a single tricritical point, signalling the encounter
of the continuous part of the frontier with the first-order one; this later aspect 
reminds topology I [cf. Fig.~\ref{fig:tricritical}(a)].  
However, as already mentioned, the tricritical point 
does not appear as the sole critical point of the phase diagram, an
effect that occurs only up to $(D/J)=0.3535$, where the fourth-order
critical point emerges, as shown in Fig.~\ref{fig:tricritical}(b).   
We have not found any qualitative change in the phase diagram of topology V by
increasing further $D/J$; in fact, comparing Figs.~\ref{fig:fifthtopology}(a) 
and (b), one notices that the
effect of increasing $D/J$ corresponds to an enlargement of phase 
${\rm \bf F_{2}}$, associated with a shift of the low-temperature
first-order critical frontier starting at $(h_{0c}/J)=(D/J) +1/4$, 
for zero temperature. 
A similar effect was verified in the case $D<0$ of fully 
correlated fields [cf. Fig.~\ref{fig:correlatedfields}(b)], 
where the zero-temperature critical point 
was shown to move according to
$(h_{0c}/J)=|D/J|+1/2$.    

\subsection{Domain-Wall Analysis for Lower-Critical Dimension}

Below we apply domain-wall arguments 
to estimate the lower-critical dimension $d_{l}$, above which an ordered state 
should occur in the corresponding nearest-neighbor version 
of the present model. Our arguments follow closely those used for the 
random-field Ising model~\cite{dotsenkobook,imryma}, which were confirmed
later by means of a rigorous proof in Ref.~\cite{imbrie}. In order to carry out 
such analysis, we rewrite~\eq{eq:hamiltonian1} as  

\begin{equation}
\label{eq:nnhamiltonian}
{\cal H}(\{h^{\tau}_{i},h^{\sigma}_{i}\}) = - J \sum_{\langle ij \rangle} \sigma_{i}\sigma_{j}
- J \sum_{\langle ij \rangle} \tau_{i}\tau_{j} - D\sum_{i=1}^{N}\tau_{i}\sigma_{i}
-\sum_{i=1}^{N}h^{\tau}_{i}\tau_{i}-\sum_{i=1}^{N}h^{\sigma}_{i}\sigma_{i}~, 
\end{equation}

\vskip \baselineskip
\noindent
where the summations $\sum_{\langle ij \rangle}$ now correspond to distinct 
nearest-neighbor pairs of spins on a regular lattice of dimension $d$.  

One should remind that the two ferromagnetic phases that appeared  
in some of the phase diagrams shown herein are characterized by a single 
order parameter, $m>0$, where 
$m = m_{\tau} = m_{\sigma}$ ($D>0$), 
or $m = m_{\tau} = -m_{\sigma}$ ($D<0$), so that these phases  
differ only by the values of the corresponding 
magnetizations, i.e.,  
${\rm \bf F_{1}}$ (higher values of $m$)  
and ${\rm \bf F_{2}}$ (lower values of $m$).
Consequently, the domain-wall analysis, which consists in estimating
energy contributions of the terms of~\eq{eq:nnhamiltonian}, is not able to 
identify multicritical points, as well as to distinguish between the two 
ferromagnetic phases; below, we consider such an analysis, which
applies to the existence of an ordered state, characterized by $m>0$, i.e., to both 
phases ${\rm \bf F_{1}}$ and ${\rm \bf F_{2}}$. 
Therefore, for testing the stability of such an ordered state, we consider the system
defined by the Hamiltonian of~\eq{eq:nnhamiltonian} in its ground state, at a 
sufficiently low temperature, and we flip the sign 
of the magnetization in a large region $R$ of the lattice, characterized by a 
typical linear size $L$. Each term in the Hamiltonian of~\eq{eq:nnhamiltonian} 
will contribute to change the ground-state energy; the ferromagnetic
interactions will produce an increase in this energy, due to the creation
of the interface,   

\begin{equation}
\label{eq:nnfincrease}
\epsilon_{J}^{\tau} \sim JL^{d-1}~; \qquad 
\epsilon_{J}^{\sigma} \sim JL^{d-1}~. 
\end{equation}

\vskip \baselineskip
\noindent
Since the fields are quenched random variables
characterized by short-range correlations, the quantities 
$\sum_{i=1}^{N}h^{\tau}_{i}\tau_{i}$ 
and $\sum_{i=1}^{N}h^{\sigma}_{i}\sigma_{i}$, for large domains, 
should approach normally distributed random variables, with 
typical values of the 
order $\pm(\overline{h_{i}^{2}} L^{d})^{1/2}=\pm h_{0}L^{d/2}$
(i.e., of the order of the width of the Gaussian distribution).
One can choose the region of flipped spins such that these contributions 
lead to a decrease of the ground-state energy, i.e.,  

\begin{equation}
\label{eq:rfdecrease}
\epsilon_{h}^{\tau} \sim - h_{0}L^{d/2}~; \qquad 
\epsilon_{h}^{\sigma} \sim - h_{0}L^{d/2}~. 
\end{equation}

\vskip \baselineskip
\noindent
Hence, if $D=0$, these two effects compete with each other,
the contributions of~\eq{eq:nnfincrease} favoring the ordered state, 
whereas those of~\eq{eq:rfdecrease} destabilizing the ferromagnetic phase. 
The change in the ground-state energy,
due to the flip in the magnetization of the region $R$, 
is estimated as 

\begin{equation}
\label{eq:d0change}
\delta = \epsilon_{J}^{\tau} + \epsilon_{J}^{\sigma} +
\epsilon_{h}^{\tau} + \epsilon_{h}^{\sigma} 
\sim 2 (JL^{d-1} - h_{0}L^{d/2})~,  
\end{equation}

\vskip \baselineskip
\noindent
which, except for the factor of 2, lead precisely to the same
contributions of the Ising ferromagnet in the presence of 
random field~\cite{dotsenkobook,imryma}.  Hence, for 
sufficiently large $L$, the ordered
state prevails for $d>2$, whereas the disordered (paramagnetic) 
phase dominates for $d<2$, from which one obtains the
lower-critical dimension $d_{l}=2$. 

The introduction of a coupling $D$ between the two Ising variables
will have no effects on the interface, whereas those 
inside the region $R$ will just affect the correlations between
the two Ising systems, i.e., they do not contribute to  
stabilize (or destabilize) the ordered state. 
Hence, the parameter $D$ in  
the Hamiltonian of~\eq{eq:nnhamiltonian} should not 
be associated with the appearance of an
ordered phase, but rather to the possible occurrence of 
multicritical behavior in the nearest-neighbor version of the model.
Therefore, one should not expect any changes in the 
lower-critical dimension $d_{l}=2$, due to the coupling between 
the two Ising variables. 
However, the corresponding energy contribution will depend on 
the dimension $d$, 
in the sense that it may change according to the state of the
system, as discussed next. 
(i) For $d<2$, where the paramagnetic state prevails, 
the contribution $-D\sum_{i=1}^{N} \tau_{i} \sigma_{i}$  
will behave like those
of~\eq{eq:rfdecrease}, yielding $\epsilon_{D} \sim \pm DL^{d/2}$;
(ii) For $d>2$, where the ordered state appears, this contribution will lead to 
$\epsilon_{D} \sim - DL^{d}$ ($D>0$), enlarging the ferromagnetic
phase, as verified in the phase diagrams presented above. 

\section{Conclusions}

We have analyzed a ferromagnetic four-state clock model in the presence 
of an anisotropy field $D$ and different conditions for random fields. 
The model was considered in the limit of infinite-range interactions, for which 
the mean-field approach becomes exact. 
By using a representation
of two Ising variables ($\{\tau_{i}\}$ and $\{\sigma_{i}\}$ for each site $i$), 
the model was expressed as two ferromagnetic 
Ising models, each with its own random field 
($\{h^{\tau}_{i}\}$  and $\{h^{\sigma}_{i}\}$, respectively). 
Moreover, in this representation, the anisotropy field leads to a coupling 
between these two variable sets, in such a way that
$D>0$ ($D<0$) favors parallel (antiparallel) alignment of the two 
Ising systems. 
We have shown that if there is no correlation between the 
random fields $\{h^{\tau}_{i}\}$  and $\{h^{\sigma}_{i}\}$, 
the Hamiltonian of the system presents a symmetry 
$D \rightarrow -D$. 
The random fields $\{h^{\tau}_{i}\}$  and $\{h^{\sigma}_{i}\}$ were 
considered as following bimodal probability distributions, in two 
extreme situations, namely, fully correlated random fields, i.e.,
$h^{\tau}_{i}=h^{\sigma}_{i}$ ($\forall i$), for which 
we have analyzed both $D>0$ and $D<0$ cases,  
and uncorrelated fields, for which we have studied  
typical values of $D>0$.  

For fully correlated fields,  
$h^{\tau}_{i}=h^{\sigma}_{i}=\pm h_{0}$, all phase diagrams 
presented the same qualitative behavior, similar to the one
of an Ising ferromagnet in the presence of a bimodal random 
field: the paramagnetic and ferromagnetic phases are separated by  
a continuous frontier at high temperatures, followed by a first-order one
for lower temperatures, with these two critical lines meeting  
at a tricritical point. 
Hence, the coupling $D$ between the two systems does not play an important
role, from the qualitative point of view. 
Quantitatively, 
the cases $D<0$ presented ferromagnetic phases that increase
significantly for sufficiently large values of $|D|$.  
  
For uncorrelated fields, since the Hamiltonian presents the 
symmetry $D \rightarrow -D$,  
we have restricted our investigation to $D>0$ only. 
This situation has shown a very rich critical behavior by varying $D$, 
with the possibility of 
two ferromagnetic phases, ${\rm \bf F_{1}}$  and ${\rm \bf F_{2}}$, 
besides the usual disordered phase {\rm \bf P}, as well as a wide variety
of critical points.   
For sufficiently small values of the coupling $D$, the phase diagram 
presents a structure typical of two independent Ising models, being 
qualitatively similar to the phase diagram of the Ising ferromagnet 
in the presence of a bimodal random field, 
characterized by a single ferromagnetic phase. 
By increasing gradually
the coupling between the two Ising systems, the additional ferromagnetic
phase emerges, with the two ferromagnetic phases, 
${\rm \bf F_{1}}$ (higher values of magnetization)  
and ${\rm \bf F_{2}}$ (lower values of magnetization), 
being separated by a first-order critical frontier that 
terminates at an ordered critical point. 

Therefore, in the case of uncorrelated fields 
we have found five well-defined types of phase diagrams, 
denominated as topologies I--V, which differ from one another by the
presence of distinct critical behavior, with
tricritical, fourth-order, ordered, triple, multicritical, and 
critical end points.
These qualitatively different types of phase diagrams correspond
to the intervals $0 < (D/J) \leq 0.3535$ (topology I), 
$0.3535 < (D/J) \leq 0.470$ (topology II), 
$ 0.472 \leq (D/J) < 0.5$ (topology III), $(D/J)=0.5$ (topology IV), and 
$(D/J) > 0.5$ (topology V). The change from topologies II and III
is very subtle from the numerical point of view, since this occurs through
the disappearance of the critical end point, giving rise to
to two other critical points, namely, a triple and a tricritical one. 
We have found that this occurs for $(D/J) = 0.471 \pm 0.001$, 
in the sense that topology II holds clearly for $(D/J)=0.470$, whereas 
topology III applies for $(D/J)=0.472$.  
From all these cases, only topology I 
typifies a well-known phase diagram, qualitatively similar to 
the Ising ferromagnet in the presence of a bimodal 
random field~\cite{aharony}.

We have carried heuristic domain-wall arguments
for estimating the lower critical dimension, above which an ordered 
state should appear in the corresponding nearest-neighbor version 
of the model. 
The study considered an ordered state, characterized by a single 
magnetization parameter, so that it applies to both 
phases ${\rm \bf F_{1}}$ and ${\rm \bf F_{2}}$. 
These arguments led to $d_{l}=2$, i.e.,
the same lower critical dimension of the Ising ferromagnet in the presence of
a random field. Our analysis indicated that the coupling $D$
does not contribute to change $d_{l}$; however, the gradual
increase of $D$ should be associated with a 
possible occurrence of multicritical behavior, as well as 
to an enlargement of the ordered phase.

From the physical point of view, the first situation considered 
herein, namely, fully correlated fields, would correspond to a situation
where in each position $i$ the fields 
$\{h^{\tau}_{i}\}$  and $\{h^{\sigma}_{i}\}$ are the same, being
associated to random effects due to the medium only. 
The second case, where in each position $i$ one has independent 
fields $\{h^{\tau}_{i}\}$  and $\{h^{\sigma}_{i}\}$, may result from 
randomnesses of the medium, in addition to other
possible effects (e.g., from the remaining spin variables),  
such as to act distinctly on the systems 
$\{\tau_{i}\}$ and $\{\sigma_{i}\}$. 
However, since the Ising model is well-known to provide a wide applicability  
in many complex systems so far, the
richness of critical behavior exhibited by the model studied herein, 
with phase diagrams presenting new and interesting
topologies, is expected to be useful for other complex phenomena, out 
of the scope of magnetism.  

\vskip 2\baselineskip

{\large\bf Acknowledgments}

\vskip \baselineskip
\noindent
The partial financial support from CNPq, 
FAPEAM-Projeto-Universal-Amazonas, 
and FAPERJ (Brazilian funding agencies) is acknowledged. 

\vskip 2\baselineskip

\end{document}